\documentclass[aps,prd,onecolumn,superscriptaddress,nofootinbib,floatfix,noshowpacs]{revtex4-2}
\pdfoutput=1
\newif\ifcomment
\usepackage{dcolumn} 
\usepackage{amsmath,amssymb,amscd,amsfonts}
\usepackage{color,hyperref,url}
\usepackage{listings}
\usepackage{slashed}
\usepackage{color}
\usepackage[pdftex]{graphicx}
\usepackage{epstopdf}
\usepackage{epsfig}
\usepackage{grffile} 
\usepackage{relsize}
\usepackage{multirow}
\usepackage{xspace}
\usepackage{float}
\usepackage{calrsfs}
\usepackage{txfonts}
\usepackage{soul}
\newcommand{\al}{\alpha}
\newcommand{\be}{\beta}
\newcommand{\ga}{\gamma}
\newcommand{\Ga}{\Gamma}
\newcommand{\de}{\delta}

\newcommand{\eps}{\epsilon}

\newcommand{\ka}{\kappa}

\newcommand{\La}{\Lambda}
\newcommand{\si}{\sigma}

\newcommand{\Om}{\Omega}

\DeclareSymbolFont{sfletters}{OML}{cmbrm}{m}{it}
\DeclareMathSymbol{\sga}{\mathord}{sfletters}{"0D}
\DeclareMathSymbol{\ssi}{\mathord}{sfletters}{"1B}
\newcommand{\beq}{\begin{equation}}
\newcommand{\eeq}{\end{equation}}
\newcommand{\ba}{\begin{array}}
\newcommand{\ea}{\end{array}}
\newcommand{\bea}{\begin{align}}
\newcommand{\eea}{\end{align}}
\newcommand{\bi}{\begin{itemize}}
\newcommand{\ei}{\end{itemize}}
\newcommand{\ben}{\begin{enumerate}}
\newcommand{\een}{\end{enumerate}}
\newcommand{\bc}{\begin{center}}
\newcommand{\ec}{\end{center}}
\newcommand{\bl}{\begin{flushleft}}
\newcommand{\el}{\end{flushleft}}
\newcommand{\br}{\begin{flushright}}
\newcommand{\er}{\end{flushright}}

\newcommand{\nn}{\nonumber \\}
\newcommand\Eqn[1]{Eq.~(\ref{#1})}  
\newcommand\Fig[1]{Fig.~(\ref{#1})} 

\newcommand{\mr}{\mathrm}
\newcommand{\mb}{\mathbf}

\newcommand{\mc}{\mathcal}
\newcommand{\mi}{\mathop{}\!i}
\newcommand{\me}{\mathop{}\!e}
\newcommand{\dif}{\mathop{}\!d}
\newcommand{\p}{\partial}

\newcommand{\tr}{\hbox{tr}}
\newcommand{\Tr}{\hbox{Tr}}

\newcommand{\GeV}{{\rm GeV}}
\newcommand{\diag}{{\rm diag}}

\renewcommand{\l}{\left}
\renewcommand{\r}{\right}

\newcommand\comment[1]{ \hbox{[{\it Comment suppressed here.}\/]} }
\newcommand\hide[1]{}
\newcommand{\skipover}[1]{}
\newcommand{\sign}{\,\mbox{sign}}

\renewcommand{\sp}{\shortparallel}
\begin{document}
\title{Dimensional reduction and the generalized pion in a magnetic field within the NJL model}
\author{Jingyi Chao}%
\email{chaojingyi@jxnu.edu.cn}
\affiliation{ 
College of Physics and Communication Electronics, Jiangxi Normal University, Nanchang, Jiangxi 330022, China}
\author{Yu-Xin Liu}\email{yxliu@pku.edu.cn}
\affiliation{Department of Physics and State Key Laboratory of Nuclear Physics and Technology, Peking University, Beijing 100871, China}
\affiliation{Collaborative Innovation Center of Quantum Matter, Beijing 100871, China}
\affiliation{Center for High Energy Physics, Peking University, Beijing 100871, China}

\date{\today}

\begin{abstract}
In this work, the mass of the neutral pion is investigated in the presence of background magnetic fields in the framework of the Nambu--Jona-Lasinio model. Taking into account the anisotropic four-fermion interactions, a tensor current arises in the magnetized QCD system, which forms an anomalous magnetic moment (AMM) coupling in the Dirac equation for the quarks. By solving the gap equations, we find that the sign of the dynamically generated AMM is opposite to the sign of the quark's charge and its magnitude is definitely smaller than the constituent mass. We construct two generalized Nambu-Goldstone pions, which emerge as combinations of the quantum fluctuations around the conventional scalar and the emergent tensor chiral condensates. We analytically demonstrate that the Goldstone nature has been spoiled by the dimensional reduction in the two-particle state and the corresponding decreasing mass of the lighter generalized pionic mode is a remnant of the infrared dynamics.
\end{abstract}


\maketitle
\section{introduction}\label{sec:intro}
Arising as a powerful probe in the study of vacuum properties and phase structure under the influence of the external magnetic field, the phase diagram of Quantum Chromodynamics (QCD) matter is extensively explored due to its relevance in the context of lattice gauge theories~\cite{Bali:2018sey,Bali:2017ian,Ding:2020jui,Ding:2020hxw}, off-central heavy-ion collisions~\cite{Skokov:2009qp,Deng:2012pc,Kharzeev:2012ph} and the merging process of neutron stars~\cite{Kaspi:2017fwg}. As one of the fundamental properties, the spectra of hadrons are used to describe the confined, chirally broken QCD phase and to construct the low-energy strong interactions. Among these quasi-particles, the pseudo-Goldstone meson plays an essential role since it is the degree of freedom carrying the chiral effective theory~\cite{Andersen:2012zc}. Unlike thermal QCD systems, the presence of the magnetic field breaks chiral symmetry from $\mr{SU}(2)$ to $\mr{U}(1)$, thus, the identified Nambu-Goldstone (NG) boson is reduced from the pseudo-scalar triplet to the individual neutral pion~\cite{Chao:2018czo}. The magnetized masses of neutral and/or charged pions have been calculated by a variety of model approaches, found in~\cite{Andersen:2012dz,Gorsky:2012ui,Avancini:2016fgq,Zhang:2016qrl,Wang:2017vtn,Liu:2018zag,Mao:2018dqe,Coppola:2019uyr,Das:2019ehv,Sheng:2020hge,Ayala:2020dxs,Xing:2021kbw}. 

In the present paper, the study of the energy dispersion of $\pi_{0}$ is motivated by the emergence of the tensor polarization of the chiral condensates in magnetized QCD matter~\cite{Gorsky:2012ui}. It is known that a spontaneous symmetry breaking appears when the Lagrangian of a system is invariant under the symmetry transformation, but the ground state is not. A more precise description is that if several ground states simultaneously break the same global symmetries, the corresponding number of NG excitations is unchanged, even though more gap equations are necessary to characterize the one-point particle state of the system. It is observed that, while the scalar vacuum expectation value (VEV) generates a dynamical fermion mass, the developed VEV of the tensor current gives rise to a dynamical anomalous magnetic moment (AMM) for the fermions~\cite{Ferrer:2008dy,bueno2012landau,Eminov:2018lse}. We stress that our purpose is not to claim that previous model calculations with fixed AMM are wrong. Indeed, the $eB$-field is not the only source responding to the spin-dependent anomaly, but also several microscopic mechanisms offer a momentum-dependent AMM for strongly interacting fermions~\cite{Bicudo:1998qb,Faccioli:2002jd,Chang:2010hb,Chang:2013nia,Gutsche:2014zua,Zhang:2017zpi}. The effects of quark AMM on the phase structure as well as mesonic properties are found in the works of~\cite{Frasca:2011zn,Fayazbakhsh:2014mca,Ferrer:2013noa,Tsue:2016age,Maruyama:2018fpl,Chaudhuri:2019lbw,Mei:2020jzn,Xu:2020yag,Aguirre:2021ljk}. 

Although the infrared dynamics of quark condensation is catalyzed by the influence of dimensional reduction in a strong magnetic field, it does not affect the motion of the neutral NG mode, as explained in~\cite{Coleman:1973ci,Gusynin:1995nb}. On the other hand, the Nambu--Jona-Lasinio (NJL) model calculation numerically shows that the AMM effect corresponds to a monotonic decrease in the spectra of the neutral pion as the strength of magnetic field grows~\cite{Xu:2020yag}, which is consistent with lattice QCD simulations~\cite{Bali:2017ian,Bali:2018sey,Ding:2020jui,Ding:2020hxw}. Once the magnitude of AMM reaches a critical value, the energy of neural pions vanishes and their condensation suggests a newly possible superfluid state in the QCD vacuum. These infrared phenomena are not in accordance with Goldstone's theorem. Also, the calculations in the low-energy effective theories observed that, taking into account the single scalar chiral condensate, the properties of the NG meson remain valid~\cite{Agasian:2001ym,Avancini:2015ady}. Hence, one shall move towards an analysis of how the unusual infrared behavior of $\pi_{0}$ is enhanced by the multiple ground states. In the present paper, we will further study the scenario of the dimensional reduction, not restricted to the one-point, but also occurring in the two-point correlators. The relation among the current quark mass, quark condensates and the mass of low-lying meson is examined, as well.

In the framework of the NJL model, mesons are treated as correlated quark-antiquark states in the random phase approximation (RPA). However, corresponding to simultaneous fluctuations of two VEVs, meson modes must be formed in terms of the excitation of $\l<\bar{\psi}\psi\r>$ and $\l<\bar{\psi}\si^{\mu\nu}\psi\r>$. While the chiral mesons have been studied as quantum fluctuations of the scalar order parameter, a quantitative representation of pseudo-scalar modes including the tensor state of $\l<\bar{\psi}\si^{\mu\nu}\psi\r>$ is still lacking. Another aim of the present article is to study the behavior of the generalized pions, contributing to the understanding of the proper degrees of freedom in the presence of external magnetic fields~\cite{Yamamoto:2007ah,Song:2017dws}. 

This paper is organized as follows: in Sec.\ref{sec:form} we introduce the NJL model Lagrangian with tensor coupling and compute the quark propagator as well as the gap equations, with two order parameters. Then, we determine the sign and the strength of $\l<\bar{\psi}\psi\r>$ and $\l<\bar{\psi}\si^{\mu\nu}\psi\r>$ for the four-fermion coupling constants $G_{S}=G_{T}$. Next, we identify the collective modes in detail in Sec.\ref{sec:pion}, where two pseudo-scalar pionic modes are presented due to the scalar-tensor mixing. We discuss the dimensional reduction appearing in the meson kernel in Sec.\ref{sec:DR} and investigate the spectrum of pions under the influence of the AMM. Finally, we discuss the results in Sec.\ref{sec:con}.
\section{Model and Formalism}\label{sec:form}
Integrating out the degrees of freedom of gluons and large quark fluctuations, whose momenta are larger than $\La_{QCD}$, the NJL model utilizes the simple four-fermion point interactions to describe spontaneous chiral symmetry breaking in QCD, which is a successful tool studied in many previous works. We will apply it to investigate the dynamics of strong interactions at low energies in a constant and homogeneous magnetic fields, without including the phenomenon of confinement, for simplicity.
\subsection{Formalism of the quark propagator}
The Lagrangian density of the NJL two-flavor model in the presence of an external magnetic field is given by
\begin{equation}
        \mc{L}=\bar{\psi}\l(\slashed{D}-m\r)\psi+G_{S}\l[\l(\bar{\psi}\psi\r)^2+\l(\bar{\psi}\mi\ga_{5}\vec{\tau}\psi\r)^2\r],
\end{equation}
where the covariant derivative $D_{\mu}=-\mi\p_{\mu}-q_{f}A_{\mu}$; $q_{f}=\diag(q_{u},q_{d})$;  $\psi$ is the quark spinor with Dirac, color and flavor indices; $A_{\mu}=\l(0,0,Bx,0\r)$ for $\mu=0,1,2,3$ and the particular constant and homogeneous magnetic field is pointing in the $x_3$-direction. As customary, we assume from the very beginning $m_{u}=m_{d}$ for the bare quark mass matrix $m$. $\vec{\tau}$ is a vector of Pauli matrices in flavor space. The conventional four-fermion scalar and pseudoscalar channels, shown in the bracket, with coupling strength $G_{S}$ are employed.

Under the influence of a uniform magnetic field, the tensor structure of the gluon propagator separates into longitudinal and transverse parts and so does the Lagrangian density of NJL, based on the effective one-gluon exchange-type interaction, given as:
\begin{align}
\mc{L}_{int}=g_{\sp}^{2}\l(\bar{\psi}\ga_{\mu}^{\sp}\psi\r)^{2}+g_{\perp}^{2}\l(\bar{\psi}\ga_{\mu}^{\perp}\psi\r)^{2}.
\end{align}
To take into account the fact that the rotation symmetry has been reduced from $O(3)$ to $O(2)$, presented in Ref.~\cite{Ferrer:2013noa}, anisotropic Fierz identities have to be applied as:
\begin{align}
        \l(\ga^{\sp}_{\mu}\r)_{il}\l(\ga_{\sp}^{\mu}\r)_{jk}&=\frac{1}{2}\l\{
\l(1\r)_{il}\l(1\r)_{jk}+\l(\mi\ga_{5}\r)_{il}\l(\mi\ga_{5}\r)_{jk}+\frac{1}{2}\l(\si_{\perp}^{\mu\nu}\r)_{il}\l(\si^{\perp}_{\mu\nu}\r)_{jk}-\l(\si^{03}_{\sp}\r)_{il}\l(\si_{03}^{\sp}\r)_{jk}
        +...\r\},\nn
        \l(\ga^{\perp}_{\mu}\r)_{il}\l(\ga_{\perp}^{\mu}\r)_{jk}&=\frac{1}{2}\l\{
\l(1\r)_{il}\l(1\r)_{jk}+\l(\mi\ga_{5}\r)_{il}\l(\mi\ga_{5}\r)_{jk}-\frac{1}{2}\l(\si_{\perp}^{\mu\nu}\r)_{il}\l(\si^{\perp}_{\mu\nu}\r)_{jk}+\l(\si^{03}_{\sp}\r)_{il}\l(\si_{03}^{\sp}\r)_{jk}+...\r\}.
\end{align}
We note here that $\sp,\perp$ carry the Lorentz indices of $(0,3)$ and $(1,2)$, respectively, regarding the direction of the magnetic field. It obviously shows that the difference between $g_{\sp,\perp}$ will manifest themselves through the frozen four-fermion interactions in the tensor channels rather than the usual interactions of scalar and pseudo-scalar couplings in the NJL model. We conclude that 
\begin{equation}\label{eqn:lag}
        \mc{L}_{\mr{int}}=G_{S}\l[\l(\bar{\psi}\psi\r)^2+\l(\bar{\psi}\mi\ga_{5}\vec{\tau}\,\psi\r)^2\r]
        +G_{T}\sum_{a=0,3}\l[\l(\bar{\psi}\si^{12}\tau_{a}\psi\r)^2+\l(\bar{\psi}\mi\ga_{5}\si^{12}\tau_a\psi\r)^2\r].
\end{equation}
$G_{T}\leq G_{S}$ since $G_{S}\sim g_{\sp}^{2}+g_{\perp}^{2}$ and $G_{T}\sim g_{\sp}^{2}-g_{\perp}^{2}$~\cite{Ferrer:2013noa}. $\tau_{a}=\l(I_{2},\vec{\tau},\r)$ and $\vec{\tau}=\tau_{1,2,3}$ are Pauli matrices. {In a magnetic environment, the vacuum state must be neutral to maintain stability. Therefore, we have omitted the non-diagonal components of the condensates in flavor space. Consequently, terms in the summation over $a$ are limited to $a = 0, 3$.} The transverse index $\si^{12}$ is selected with respect to the magnetic field in the $x_3$-direction.
{As discussed in Ref.~\cite{Ferrer:2013noa}, the positive definiteness of $G_{T}$ arises from the dominance of longitudinal contributions from one-gluon exchange over transverse ones due to the emergence of a dimension reduction effect caused by the magnetic field; i.e., $G_{S}\sim G_{T}$ for negligible $g_{\perp}$ in strong magnetic fields.} 

{In the presence of an external magnetic field, the $\operatorname{SU}(2)\times \operatorname{SU}(2)$ chiral symmetry of the two-flavor NJL model is explicitly broken down to $\operatorname{U}(1)_{I{3}}\times \operatorname{U}(1)_{AI{3}}$. Both a chiral condensate and a tensor condensate break the invariant Lagrangian to $\operatorname{U}(1)_{L+R}$. The chiral condensate creates a mass gap for the quarks. The tensor condensate generates an anomalous magnetic moment for the quarks. We examine the phase structure of the model based on these two condensates and their dependence on the quark charge.} Written as: 
\begin{align}\label{eqn:mean}
        \Sigma_f=-G_{S}\l<\bar{\psi}\psi\r>_f\,,\quad 
        \ka_f=-G_{T}\l<\bar{\psi}\si^{12}\psi\r>_f\,, 
\end{align}
{for $f=u,d$. We adopt the notation that $\Sigma =\Sigma_u+\Sigma_d$ and assume that the chiral condensate $\Sigma_u=\Sigma_d$ for maximal flavor symmetry.} As we mentioned before, a non-trivial coupling constant $\ka_f$ of the anomalous magnetic moment is produced through several microscopic mechanisms. The coefficient of quark AMM is not uniquely adopted in many previous works~\cite{Frasca:2011zn,Fayazbakhsh:2014mca,Ferrer:2013noa,Tsue:2016age,Maruyama:2018fpl,Chaudhuri:2019lbw,Mei:2020jzn,Xu:2020yag,Aguirre:2021ljk}, which is proportional to either $q_{f}$, or $q_{f}^{2}$, or charge independent if it is created via a compensation of the color-AMM. 
The main point in the present investigation is that we will dynamically determine and extract it from the gap equations in the following. %

Performing the Hubbard-Stratanovich transformation in the Lagrangian density of \Eqn{eqn:lag} and plugging into the \Eqn{eqn:mean}, we continue to derive the magnetized quark propagator with the AMM coupling. 
Hence, the fermionic Lagrangian density in the mean-field approximation is rewritten as:
\begin{equation}
        \mc{L}_{\mr{eff}}=\bar{\psi}\l(\slashed{D}-M+\vec{\kappa}\cdot\tau_a\, \si^{\mu\nu}\hat{F}_{\mu\nu}\r)\psi,
\end{equation}
where $M=\Sigma +m$, $\si^{\mu\nu}=\mi\l[\ga^{\mu},\ga^{\nu}\r]/2$.
{By summing over $a=0,3$, we obtain the two-vector $\vec{\kappa}=\frac{1}{2}\l(\kappa_u+\kappa_d,\kappa_u-\kappa_d\r)$ that represents $\vec{\kappa}\cdot\tau_a=\diag(\ka_u,\ka_d)$ in flavor space.} 
$\hat{F}_{\mu\nu}=F_{\mu\nu}/||F||$ is the dimensionless electromagnetic (EM) tensor. Note here that we let the energy scale of $\ka_f$ be the same as the mass, which was scaled to a dimensionless quantity in some works.

{To study the behavior of $\vec{\kappa}$, we examine its dependence on the quark charge in the one-flavor model and neglect the vector symbol. It is important to note that $\kappa_f$ is a flavour dependent parameter in both the one and two flavor models. This is pointed out throughout the manuscript.} Following Schwinger's proper time method, we obtain the quark propagator as:
\begin{align}
G&=\frac{1}{\slashed{D}-M+\ka \, \ssi\hat{\mathsf{F}}}
=\frac{\slashed{D}+M+\ka \, \ssi\hat{\mathsf{F}}}{\slashed{{D}}^{2}-M^{2}+\ka^{2} +\slashed{\Om}}=\mi\l(\slashed{D}+M+\ka \, \ssi\hat{\mathsf{F}}\r)\int\dif s\me^{\mi s\l(\slashed{D}^{2}-M^2+\ka^{2} +\slashed{\Om}\r)},
\end{align}
where $\slashed{D}^{2}=D^{2}-{q_{f}\ssi\mathsf{F}}/{2}$. The formula is given in matrix notation, e.g. $F_{\mu\nu}=(\mathsf{F})_{\mu\nu}$, $\ssi\mathsf{F}=\si^{\mu\nu}F_{\mu\nu}$. The additional AMM term gives
\beq
\slashed{\Om}=-2\mi\ka \, \l(\ga^3\ga^5 \p_{0}-\ga^0\ga^5 \p_{3}\r).
\eeq
$\slashed{\Om}$ commutes with $\slashed{D}^{2}$ with $\perp=1,2$, since $\l[(\si^{\mu\nu})_{\perp},\ga^{(0,3)}\ga^{5}\r]=0$. At this point, it allows for an expansion of the exponential~\cite{Dittrich:2000zu}
\beq
\me^{\mi\slashed{\Om}s}=\cosh\l(\mi\Om s\r)+\frac{\slashed{\Om}}{\Om}\sinh\l(\mi\Om s\r),
\eeq
with constant matrix $\Om=||\slashed{\Om}||=2|\ka|\sqrt{-\p^{2}_{0}+\p^{2}_{3}}$. Finally, the Green’s function takes the form:
\begin{align}\label{eqn:G}
&G(x,y)=\frac{\phi(x,y)}{4\pi^{2}}\sum_{\pm}\int\frac{\dif s}{s^2}\me^{\mi s\l(\Pi^{2}-M^2+\ka^{2}\pm\Om\r)-L(s)}\l[\frac{1}{2}\ga^{\mu}\l(\mathsf{f}(s)+q_{f}\mathsf{F}\r)_{\mu\nu}(x-y)^{\nu}+M+\ka \, \ssi\hat{\mathsf{F}}\r]\l[1\pm\frac{\slashed{\Om}}{\Om}\r],
 \end{align}
where $\phi(x,y)$ is the well-known phase factor~\cite{Schwinger:1951nm,Miransky:2015ava} and 
\begin{align}
&\Pi^{2}=\frac{1}{4s}(x-y)\mathsf{f}(s)(x-y)+\frac{q_{f}\ssi\mathsf{F}}{2};
\quad  
\mathsf{f}(s)=q_{f}\mathsf{F}\coth\l(q_{f}\mathsf{F}s\r);\quad
L(s)=\frac{1}{2}\tr\ln\frac{\sinh\l(q_{f}\mathsf{F}s\r)}{q_{f}\mathsf{F}s}.
\end{align}
The position dependence of $G(x,y)$ has been attributed to the Schwinger phase factor $\phi(x,y)$ and the left term in \Eqn{eqn:G} is translation invariant. It is convenient to transform it to momentum space and further decompose over the Landau pole, representing it as
\begin{align}
 \tilde{G}(q_{f},k)=\exp\l[-\frac{ k_{\perp}^2}{|q_{f}|B}\r]\sum_{\pm}\sum_{n=0}^{\infty}(-1)^{n}
 \frac{\slashed{D}_{n}(q_{f}B,k)\,\La_{\pm}}{k_{\sp}^{2}-2n|q_{f}|B-M^{2}+\ka^{2}\pm 2|\ka k_{\sp}|}
\end{align}
with
\beq
\La_{\pm}=\frac{1}{2}\pm\frac{\ga^3\ga^5 k_{0}-\ga^0\ga^5 k_{3}}{2|k_{\sp}|}\sign(\ka),
\eeq
and
\begin{align}
\slashed{D}_{n}(q_{f},k)=\l(\slashed{k}_{\sp}+M+\kappa\,\si\hat{F}\r)\l[P_{-}L_{n}\l(2z_{f}\r)-P_{+}L_{n-1}\l(2z_{f}\r)\r]+4\slashed{k}_{\perp}L_{n-1}^{1}\l(2z_{f}\r).
\end{align}
We note here that $P_{\pm}=1\pm\mi\ga^{1}\ga^{2}\sign(q_{f})$, $z_{f}= k_{\perp}^{2}/\l(|q_{f}|B\r)$, $k_{\sp}=\l(k_{0},k_{3}\r)$, $k_{\perp}=\l(K_{22},K_{12}\r)$, $\ga_{\sp}=\l(\ga_{0},\ga_{3}\r)$ and $\ga_{\perp}=\l(\ga_{1},\ga_{2}\r)$ as usual. 

{We derive the gap equations with respect to the order parameters $\Sigma$ and $\kappa$ for a fixed electrical charge $q_f$,} the dynamical solutions are given as:
\begin{align}
        \frac{M-m}{2\mi G_{S}}&=N_{c}\,\tr\,G;\label{eqn:gap1}\\
        \frac{\ka}{2\mi G_{T}}&=N_{c}\,\tr\l[\si^{12}G\r],\label{eqn:gap2}
\end{align}
The notation of $(\tr)$ runs in Dirac and coordinate spaces, one has
\beq
\tr\,G(k)=\sum_{\pm}\sum_{n=0}^{\infty}(-1)^{n}\int\frac{\dif^{4}k}{8\pi^{4}}\me^{-z_{f}}\frac{M\l(L_{n}-L_{n-1}\r)-\xi|\ka|\l(L_{n}+L_{n-1}\r)\mp\xi\,|k_{\sp}|\l(L_{n}+L_{n-1}\r)}{\l(|k_{\sp}|\pm|\ka|\r)^{2}-M_{n}^{2}},
\eeq
where $|k_{\sp}|=\sqrt{k_{0}^{2}-k_{3}^{2}}$, $M_{n}^{2}=M^{2}+2n|q_{f}|B$ and $\xi=\sign(\ka\cdot q_{f})$. Since the role of $\ka$'s sign has been attributed to a function of $\xi$, from now on, we abbreviate $|\ka|$ to $\ka$. 
It is known that $L_{n-1}$ vanishes for $n=0$ and
\beq
\int\frac{\dif^{2}k_{\perp}}{4\pi^{2}}\me^{-z_{f}}(-1)^{n}L_{n}=\frac{|q_{f}|B}{4\pi};\qquad
\int\frac{\dif^{2}k_{\perp}}{4\pi^{2}}\me^{-z_{f}}(-1)^{n}L_{n-1}=-\frac{|q_{f}|B}{4\pi}.
\eeq
After integrating over the transverse momentum space, one has
\begin{align}\label{eqn:traceG}
        \tr\,G(k)=\frac{|q_{f}|B}{8\pi^3}\sum_{\pm}\int
        \dif^{2}k_{\sp}\l\{
        \frac{M-\xi\kappa\mp\xi\,|k_{\sp}|}{\l(|k_{\sp}|\pm\ka\r)^{2}-M^{2}}
+\sum_{n=1}^{\infty}\frac{2M}{\l(|k_{\sp}|\pm\ka\r)^{2}-M_{n}^{2}}\r\}.
\end{align}

\subsection{Sign of the AMM}

While $\ka<M$, the double degenerate roots of the denominators $\l(|k_{\sp}|\pm\ka\r)^{2}-M_{n}^{2}$ are $k_{0}=a\sqrt{k_{3}^{2}+\l(M_{n}\mp\ka\r)^{2}}-\mi a\eps$ for $a=\pm1$. For the real roots of $f(x)$ located at $x_0$, one has to apply the Jacobian feature of the Dirac-delta function
\beq
\de\l[f(x)\r]=\frac{\de(x-x_0)}{|f'(x_0)|}.
\eeq
Then, we close the contour of the semicircle in the upper half plane to complete the integral of \Eqn{eqn:traceG} with respect to $k_{0}$, shown as 
\begin{align}\label{eqn:trG}
        \tr\, G(k)=&-\mi\frac{|q_{f}|B}{4\pi^2}\int
        \dif k_{3}\l\{
        \frac{M+\xi\ka}{\sqrt{k_{3}^{2}+\l(M+\xi\ka\r)^{2}}}
        +\sum_{\pm}\sum_{n=1}^{\infty}\frac{M}{M_{n}}\frac{M_{n}\pm\ka}{\sqrt{k_{3}^{2}+\l(M_{n}\pm\ka\r)^{2}}}\r\}.
\end{align}
It is noticed that only one of the Zeeman splitting states, $\La_{\pm}$, has been survived in the Lowest Landau Level (LLL), which is not determined by the charge of quarks but with the product of $\xi$ instead. {From the right-hand side of the above equation, we can see that $(\mi\,\tr\, G)$ is positive when $M > \kappa$.}

Similarly, we obtain the related trace of $\ka$ as
\begin{align}
        \tr\l[\si^{12}G(k)\r]&
=\sum_{\pm}\sum_{n=0}^{\infty}(-1)^{n}\int\frac{\dif^{4}k}{8\pi^{4}}\me^{-z_{f}}\frac{\l(\ka\pm|k_{\sp}|\r)\l(L_{n}-L_{n-1}\r)\sign\l(\ka\r)-M\l(L_{n}+L_{n-1}\r)\sign\l(q_{f}\r)}{\l(|k_{\sp}|\pm|\ka|\r)^{2}-M_{n}^{2}}\nn
        &=\frac{|q_{f}|B}{8\pi^3}\sum_{\pm}\int
        \dif^{2}k_{\sp}\l\{
        \frac{\l(\kappa\pm|k_{\sp}|\r)\sign\l(\ka\r)-M\sign\l(q_{f}\r)}{\l(|k_{\sp}|\pm\ka\r)^{2}-M^{2}}
 +\sum_{n=1}^{\infty}\frac{2\l(\kappa\pm|k_{\sp}|\r)\sign\l(\ka\r)}{\l(|k_{\sp}|\pm\ka\r)^{2}-M_{n}^{2}}\r\}.
\end{align}
{Adopting the sign function $\sign(x)$, which satisfies $\sign(x) \cdot x = \operatorname{Abs}(x)$ for $x \neq 0$, we can determine the sign of $\kappa$ as follows:}
\begin{align}\label{eqn:abska:gap}
        \sign\l(\ka\r)\tr\l[\mi\si^{12}G(k)\r]
&=\frac{|q_{f}|B}{8\pi^3}\sum_{\pm}\int
        \dif^{2}k_{\sp}\l\{
        \frac{\l(\kappa\pm|k_{\sp}|\r)-M\sign\l(\xi\r)}{\l(|k_{\sp}|\pm\ka\r)^{2}-M^{2}}
 +\sum_{n=1}^{\infty}\frac{2\l(\kappa\pm|k_{\sp}|\r)}{\l(|k_{\sp}|\pm\ka\r)^{2}-M_{n}^{2}}\r\}\nn
        &=\frac{|q_{f}|B}{4\pi^2}\int
        \dif k_{3}\l\{\frac{-\ka-\xi M}{\sqrt{k_{3}^{2}+\l(M+\xi\ka\r)^{2}}}
        +\sum_{\pm}\sum_{n=1}^{\infty}\frac{\mp M_{n}-\ka}{\sqrt{k_{3}^{2}+\l(M_{n}\pm\ka\r)^{2}}}\r\}.
\end{align}
It is observed that the l.h.s.~of \Eqn{eqn:abska:gap} is positive definite, thus, one needs $\xi=-1$ to get a nontrivial solution of $\ka$. {Using the LLL approximation and taking the chiral limit $m \to 0$, we recover the result that $M/\kappa = G_S/G_T$~\cite{Ferrer:2013noa}. Moreover, the contribution from finite Landau levels (i.e., the second term in the above bracket) is negative since the absolute values coming from $(M_n + \kappa)$ are larger than those from $(M_n - \kappa)$. }

In the second case of $M_{l}^{2}<\ka^{2}<M^{2}_{l+1}$, there is no root in the denominators $\l(|k_{\sp}|+\ka\r)^{2}-M_{n}^{2}$ for $n\leq l$, on the contrary, the root is four-fold degenerate in the term of $\l(|k_{\sp}|-\ka\r)^{2}-M_{n}^{2}$, known as $k_{0}=\pm\sqrt{k_{3}^{2}+\l(\ka+a M_{n}\r)^{2}}\mp\mi a\eps$ with $a=\pm1$. Hence, two poles, $-\sqrt{k_{3}^{2}+\l(\ka+ M_{n}\r)^{2}}$ and $\sqrt{k_{3}^{2}+\l(\ka- M_{n}\r)^{2}}$, will contribute while taking the Cauchy integral in the upper half plane~\cite{Pisarski:2020dnx}. For $n\geq l+1$, the root behaviors of $\l(|k_{\sp}|\pm\ka\r)^{2}-M_{n}^{2}$ reduce to double degenerate states as usual. Without loss of generality, let $l=1$, $\sign\l(\xi\r)=-1$, and then,
\begin{align}
        \mathbb{S}_{\si}&=\tr\,G(k)
        \nn&
        =-\mi\frac{|q_{f}|B}{4\pi^2}\int\dif k_{3}\l\{
        \frac{\ka-M}{\sqrt{k_{3}^{2}+\l(\ka-M\r)^{2}}}
        +\sum_{\pm}
        \frac{M}{M_{1}}\frac{\ka\pm M_{1}}{\sqrt{k_{3}^{2}+\l(\ka\pm M_{1}\r)^{2}}}
        +\sum_{\pm}\sum_{n=2}^{\infty}\frac{M}{M_{n}}\frac{M_{n}\pm\ka}{\sqrt{k_{3}^{2}+\l(M_{n}\pm\ka\r)^{2}}}\r\},
\end{align}
\begin{align}\label{eqn:bigka:gap}
        \mathbb{S}_{\ka}&=\sign\l(\ka\r)\tr\l[\si^{12}G(k)\r]\nn
        &=-\mi\frac{|q_{f}|B}{4\pi^2}\int
        \dif k_{3}\l\{\frac{\ka-M}{\sqrt{k_{3}^{2}+\l(\ka-M\r)^{2}}}
        +\sum_{\pm}
        \frac{\mp\ka-M_{1}}{\sqrt{k_{3}^{2}+\l(\ka\pm M_{1}\r)^{2}}}
        +\sum_{\pm}\sum_{n=2}^{\infty}\frac{\mp M_{n}-\ka}{\sqrt{k_{3}^{2}+\l(M_{n}\pm\ka\r)^{2}}}\r\}.
\end{align}
Since
\begin{align}
       f(x)=\int_{-\La}^{\La}\dif k_{3}\frac{x}{\sqrt{k_{3}^2+x^{2}}}=x\log\frac{\sqrt{\La^{2}+x^{2}}+\La}{\sqrt{\La^{2}+x^{2}}-\La},
\end{align}
one notices that $f(x)$ is increasing as $x$ is growing and then $\mathbb{S}_{\ka}<\mathbb{S}_{\si}$. Therefore, it means that {\it no solution exists in the second case} for $G_{S}\geq G_{T}$ after comparing between the dynamical solutions of $M$ and $\ka$. The same conclusion can be drawn if we let $\xi=1$. 

Taking into account to our earlier-reached conclusion in the first case of $\ka<M$, it requires that $\xi=-1$. 
{Returning to a two-flavor quark state consisting of up and down quarks, we find that the allowed non-trivial solution has a generalized form of $\diag\l(-\ka_u \operatorname{sign}(q_u),\, -\ka_d \operatorname{sign}(q_d)\r)$, which is equivalent to $\vec{\kappa}\cdot\tau_a$, where $\kappa_u$ and $\kappa_d$ are defined as non-negative. Here we restrict ourselves to keeping the maximum chiral condensate; namely, $\kappa_u=\kappa_d=\ka$. With $q_f=\diag(e/3,-2e/3)$, we obtain the solution as $\vec{\kappa}=\l(0,-\ka\r)$ where the only non-zero component of the Pauli matrix is $\tau_3$.} 
It also allows us to convert the gap equation (\ref{eqn:gap2}) to a absolutely definite $\ka$, presented as
\begin{equation}
        \frac{|\ka|}{2\mi G_{T}}=-\Tr\l[\tau_{3}\si^{12}G(k)\r], \label{eqn:gap}
\end{equation}
where the notation of capital trace ($\,\Tr=N_{c}\sum_{q_f=u,d}\tr\,$) runs in color, flavor, Dirac and coordinate spaces. 
Hence, the first conclusion in the present work is that the sign of the dynamically generated AMM is not arbitrary, and it must be opposite to the sign of the quark charge. Besides, the magnitude of $\kappa$ is smaller than the dynamically generated quark mass, if no other sources are taken into account in the current two-flavor NJL model approach.
\section{Aspects of the Generalized Pions}\label{sec:pion}
In this Section, we show the essential features of the mixing of the generalized pseudo-Goldstone modes in the description of the NJL model.

\begin{figure}[ht]
\includegraphics[width=0.8\textwidth]{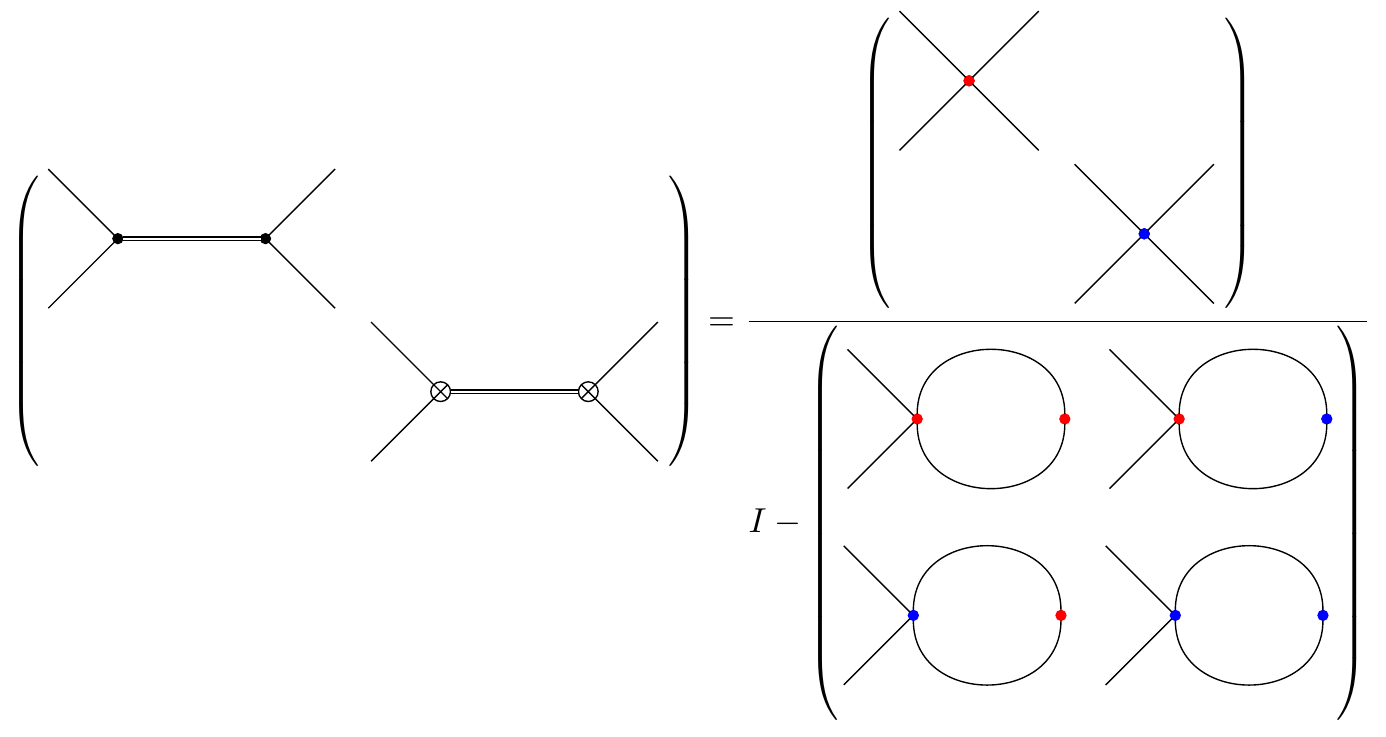}
\caption{Characteristic diagrams corresponding to the quark-antiquark T-matrix with two ground states.
{The red dot vertex denotes $(\bar{\psi}\mi\gamma_5 \tau_3 \psi)^2$ and the blue dot vertex denotes $(\bar{\psi}\mi \gamma_5 \sigma ^ {12} \tau_3 \psi)^2$. The four-fermion coupling $(\bar{\psi}\mi \gamma_5 \tau_3 \psi)(\bar{\psi}\mi \gamma_5 \sigma^{12} \tau_3 \psi)$ and its conjugate term are absent in the leading order (i.e., the off-diagonal elements in the numerator of the right above corner).}
}\label{fig:rpa}
\end{figure}

Following the discussions of Refs.~\cite{Yamamoto:2007ah,Song:2017dws}, while the low-energy effective Lagrangian is written in terms of the two order parameter fields, its associated collective modes are presented by two condensates as a model-independent consequence.
{While scalar and tensor condensates break chiral symmetries}, in this context, it is also instructive to describe the pion as the generalized one, which is the excitation of the simultaneous fluctuations on account of $\l<\bar{\psi}\psi\r>$ and $\l<\bar{\psi}\si^{12}\psi\r>$. Iterating the vertex of the four-fermion interactions, the meson is defined as the solution to the Bethe-Salpeter equation for the bound states. The equation reads
\begin{align}
        1-2G_{S}\Pi_{\operatorname{ps}}\l(m_{\pi}^{2}\r)=0,
\end{align}
where $\Pi_{\operatorname{ps}}$ is the ordinary quark-antiquark polarization tensor for (pseudo)-scalar. While the tensor condensate exists, another meson correlation of arises through the (pseudo)-tensor channels~\cite{Yamamoto:2007ah,Song:2017dws}. {As a result,  we have two sets of pion triplets}. Besides, these two sets are mixed by the interaction, seen the the off-diagonal Feynman diagram in \Fig{fig:rpa}. {Here, the lighter neutral pion $\pi_{0}$ remains as the pseudo NG mode of spontaneous chiral symmetry breaking}.

When the scalar-tensor mixing vanishes, its properties can be calculated one-by-one, which is exactly the situation of $\kappa=0$. However, considerable differences are caused when nonvanishing and one has to calculate the two-by-two matrix of the polarization tensor to correctly describe the NG modes. 

Now, the NG mesons are superpositions of ordinary quark-antiquark fluctuations $\Pi^{SS}$, plus the fluctuations of tensor quark-antiquark $\Pi^{TT}$~\cite{Xing:2021kbw}. {Note that $S$ and $T$ label the Lorentz index}. As demonstrated by \Fig{fig:rpa}, in terms of two fields $(\pi)\equiv\l(\pi,\tilde{\pi}\r)^{T}$, the T-matrix in the random phase approximation is extended as
\begin{align}\label{eqn:T}
\frac{1}{\mi}\Pi_{\operatorname{ps}}=\frac{1}{\mi} \begin{pmatrix}\:
\Pi^{SS} & \Pi^{ST} \\
\Pi^{TS} & \Pi^{TT} 
\:\end{pmatrix}
\end{align}
where
\begin{align}
        \frac{1}{\mi}\Pi^{AB}_\al&=-N_{c}\sum_{q_{f}}\tr\l[\mi G(p)\mi\ga_{5}\Ga_\al^{A}\mi G(q)\mi\ga_{5}\Ga_\al^{B}\r],
\end{align}
for $A,B=S$ and $T$.
Here, $G$ is the fermion propagator and $\Ga^{(A,B)}_{\al}=\l(I_{4}\tau_{\al},\si^{12}\tau_{\al}^{*}\r)$, respectively. {For $\alpha = 0,\pm1,3$, the quark bubble corresponds to the meson polarization function of $\sigma$ and pion triplet ($\pi_{\pm}, \pi_{0}$). As we demonstrate below, mixing makes one of the two pions heavier while the other becomes lighter. We write down} the mass spectra of $\hat{\pi}_{\al}$ and $\bar{\pi}_{\al}$, which are described as the two eigenvalues corresponding to the transformation of,
\begin{align}
        \begin{pmatrix}
                \pi\\
                \tilde{\pi}
        \end{pmatrix}
        =\mc{F}
        \begin{pmatrix}
                \hat{\pi}\\
                \bar{\pi}
        \end{pmatrix},
\end{align}
where the rotation matrix $\mc{F}^{-1}$ is applied to diagonalize the T-matrix of $\Pi_{\operatorname{ps}}$.
\section{Dimensional Reduction in the Two-particle State}\label{sec:DR}
In this section, we complete the calculation of the T-matrix of \Eqn{eqn:T} and demonstrate the dimensional reduction in the NG meson kernel. {In this section, we computed the neutral pion and $\tau_\alpha = \tau_3$. The subscript $\alpha$ of $\Pi$ is omitted.}

\subsection{Polarization tensor}
Since $\xi=-1$, one has
\begin{subequations}
\begin{align}
        \Pi^{SS}
        &=N_{c}\sum_{q_{f},\pm}\frac{|q_{f}|B}{16\pi^3}\int\dif^{2}k_{\sp}\Bigg\{
        \l(1-\frac{p_{0}q_{0}-p_{3}q_{3}}{|p_{\sp}q_{\sp}|}\r)
        \frac{\l(M+\kappa\pm|p_{\sp}|\r)\l(M+\kappa\pm|q_{\sp}|\r)}{\l[\l(|p_{\sp}|\pm\kappa\r)^{2}-M^{2}\r]\l[\l(|q_{\sp}|\pm\kappa\r)^{2}-M^{2}\r]}\label{eqn:line-1}\\
        &+\l(1+\frac{p_{0}q_{0}-p_{3}q_{3}}{|p_{\sp}q_{\sp}|}\r)
        \frac{\l(M+\kappa\pm|p_{\sp}|\r)\l(M+\kappa\mp|q_{\sp}|\r)}{\l[\l(|p_{\sp}|\pm\kappa\r)^{2}-M^{2}\r]\l[\l(|q_{\sp}|\mp\kappa\r)^{2}-M^{2}\r]}\label{eqn:line-3}\\
        &+2\sum_{n=1}^{\infty}\l(1-\frac{p_{0}q_{0}-p_{3}q_{3}}{|p_{\sp}q_{\sp}|}\r)
        \frac{M_{n}^{2}+\l(|p_{\sp}|\pm\kappa\r)\l(|q_{\sp}|\pm\kappa\r)}{\l[\l(|p_{\sp}|\pm\kappa\r)^{2}-M_{n}^{2}\r]\l[\l(|q_{\sp}|\pm\kappa\r)^{2}-M_{n}^{2}\r]}\label{eqn:line-5}\\
        &+2\sum_{n=1}^{\infty}\l(1+\frac{p_{0}q_{0}-p_{3}q_{3}}{|p_{\sp}q_{\sp}|}\r)
        \frac{M_{n}^{2}-\l(|p_{\sp}|\pm\kappa\r)\l(|q_{\sp}|\mp\kappa\r)}{\l[\l(|p_{\sp}|\pm\kappa\r)^{2}-M_{n}^{2}\r]\l[\l(|q_{\sp}|\mp\kappa\r)^{2}-M_{n}^{2}\r]}\Bigg\},\label{eqn:line-7}
\end{align}\label{eqn:all-lines}
\end{subequations}
where $p=k+\frac{M_{\pi}}{2}$, $q=k-\frac{M_{\pi}}{2}$ and $M_{\pi}=(m_{\pi},0,0,0)$ in the center-of-mass frame. The influence of the flavor matrix $\tau_{3}$ is trivial in constructing the neutral pion. After a tedious but straightforward calculation, one gets
\begin{align}\label{eqn:piSS}
        \frac{1}{\mi}\Pi^{SS}
        &=
        \frac{I_{1}}{M-\kappa}+\frac{I_{2}}{M}-m_{\pi}^{2}\l<J\r>_{0}-m_{\pi}^{2}\l<K_{11}\r>_{n}
\end{align}     
where
\begin{align}
        I_{1}&=N_{c}\sum_{q_{f}}
        \frac{|q_{f}|B}{8\pi^3}\int\frac{\dif^{2}k_{\sp}}{|k_{\sp}|-M+\kappa};
        \quad
        I_{2}=2MN_{c}\sum_{q_{f}}
        \frac{|q_{f}|B}{8\pi^3}\sum_{\pm}\sum_{n=1}^{\infty}\int\frac{\dif^{2}k_{\sp}}{\l(|k_{\sp}|\pm\kappa\r)^{2}-M_{n}^{2}}.
\end{align}
The term of $I_{1}/\l(M-\kappa\r)$ is derived from \Eqn{eqn:LLL}, and 
further explanation is provided below. Here we have introduced the brackets as: 
\begin{align}
        \l<X(k_{\sp},m_{\pi})\r>_{0}&=N_{c}\sum_{q_{f}}\frac{|q_{f}|B}{16\pi^3}\int\frac{\dif^{2}k_{\sp}}{2|p_{\sp}q_{\sp}|}X(k_{\sp},m_{\pi});\nn
        \l<X(n,k_{\sp},m_{\pi})\r>_{n}&=N_{c}\sum_{q_{f}}\frac{|q_{f}|B}{16\pi^3}\sum_{\pm}\sum_{n=1}^{\infty}\int\frac{\dif^{2}k_{\sp}}{2|p_{\sp}q_{\sp}|}X(n,k_{\sp},m_{\pi}),
\end{align}
to denote the associated summation and integration in $k$-space. The detail forms of $J,K_{11}$ are shown in the Appendix.

Before doing a full computation, we show the resulting expressions and conclusions for zero $\kappa$ and zero $M$. Firstly, it is easily found that $\Tr\,G\sim \l(I_{1}+I_{2}\r)$, read off \Eqn{eqn:traceG}. Here, the term of $I_{1}$ is rewritten as,
\begin{equation}
        I_{1}\l(M,\kappa\r)=2\l(M-\kappa\r)N_{c}\sum_{q_{f}}
        \frac{|q_{f}|B}{8\pi^3}\int\frac{\dif^{D-2}k_{\sp}}{k_{\sp}^2-\l(M-\kappa\r)^2},
\end{equation}
which is the pole contribution in the fermion propagator from the famous lowest-landau-level. In $2+1$ dimensions, $I_{1}$ remains finite in the limit of $(M,\,\kappa)\to 0$. This suggests that the infrared dynamics appearing in the one-point correlator is in response to enhancing fermion masses by a strong magnetic field in $3+1$ dimensions. It also means that the motion of charged fermions is restricted in the lower dimensions, i.e., $D\to D-2$.
Moreover, $\pi_{0}$ is determined by the original polarization tensor $\Pi^{SS}$ in \Eqn{eqn:piSS} while $\kappa=0$. Continuing to solve the Bethe-Salpeter equation, one observes that $\pi_{0}$ becomes massless in the chiral limit, as presented in Ref~\cite{Avancini:2015ady}. We emphasize here that it is consistent with the conclusions drawn in Ref.~\cite{Gusynin:1995nb}, that the dimensional reduction catalyzes the condensate $\l<\bar{\psi}\psi\r>$, but does not affect the dynamics of the neutral meson excitation since it is the same infrared term that arose in the one- and two-point correlators. We can attribute the IR-divergent $I_{1}$ in meson kernels to the dynamical quark mass $M$. Therefore, the Goldstone nature of $\pi_{0}$ is preserved and the propagators of neutral hadrons are well-behaved in $D$-dimension.

Returning to our procedure with non-vanishing $\kappa$, the other analogous expressions of polarization tensors are presented as:
\begin{align}
        \Pi^{TT}
        &=
        \frac{I_{1}}{M-\kappa}-\frac{I_{2}}{M}-m_{\pi}^{2}\l<J\r>_{0}-m_{\pi}^{2}\l<K_{22}\r>_{n},
\end{align}
and
\begin{align}
        \Pi^{ST}=\Pi^{TS}=
        \frac{I_{1}}{M-\kappa}-m_{\pi}^{2}\l<J\r>_{0}-m_{\pi}^{2}\l<K_{12}\r>_{n}.
\end{align}
Again, the detailed forms of $K_{12,22}$ can be found in the Appendix.

\subsection{Matrix of the meson kernel}
Regarding the fluctuations of two mean fields, the BS equation of meson modes converts to
\begin{align}\label{eqn:matrix}
 \begin{pmatrix}\:
1-2g\Pi^{SS} & -2g\Pi^{ST} \\
-2g\Pi^{TS} & 1-2g\Pi^{TT} 
\:\end{pmatrix}
=\mb{A}+m_{\pi}^2\,\mb{B}
\end{align}
where
 \begin{align}
\mb{A}= 1-2g\Pi_{\operatorname{ps}}\big|_{m_{\pi}^2=0}
= \begin{pmatrix}\:
        \eta&0\\
        0&2-\eta
\:\end{pmatrix}-\frac{2\mi gI_{1}}{M}\frac{1}{1-\zeta}
 \begin{pmatrix}\:
\zeta  & 1 \\
1 & 2-\zeta
\:\end{pmatrix},
\end{align}
\begin{align}
        \mb{B}=2\mi g
 \begin{pmatrix}\:
        \l<J\r>_{0}+\l<K_{11}\r>_{n}&\l<J\r>_{0}+\l<K_{12}\r>_{n}\\
        \l<J\r>_{0}+\l<K_{12}\r>_{n}&\l<J\r>_{0}+\l<K_{22}\r>_{n}
\:\end{pmatrix},
\end{align}
$G_{S}=G_{T}=g$, $\eta=\frac{m}{M}$ and $\zeta=\frac{\kappa}{M}$. 

In the polarization tensor $\Pi_{\operatorname{ps}}$, all elements acquire the pole contribution of the LLL, seen \Eqn{eqn:LLL}, which take the form
\begin{align}\label{eqn:tildeI1}
        \tilde{I}_{1}\l(M,\kappa\r)&=N_{c}\sum_{q_{f}}
        \frac{|q_{f}|B}{8\pi^3}\int\frac{\dif^{D-2}k_{\sp}}{|k_{\sp}|\l(|k_{\sp}|-M+\kappa\r)}\nn
        &=N_{c}\sum_{q_{f}}
        \frac{|q_{f}|B}{4\pi^3}\int\frac{\dif^{D-2}k_{\sp}}{k_{\sp}^2-\l(M-\kappa\r)^2}.
\end{align}
For $D=2+1$, $\tilde{I}_{1}$ is governed by the diverging integrand $\dif k/k^2$ in the limits of $(M,\,\kappa)\to 0$, while it retains logarithmic singularity in IR limits for $D=3+1$, demonstrating the emergent dimensional reduction in the neutral mesonic excitations. Under the LLL approximation, it is observed that $M=\kappa+m$ and $I_{1}\sim M\sim m\ln m^2$. Thus, the infrared dynamics has a strong
hierarchy of meson and quark sectors, where $\tilde{I}_{1}\sim\ln m^2$ differs from of $I_{1}$ with $m\to 0$. The NG bosons are formed in the infrared region, which cannot be washed out by the dynamical quark mass and results in a remarkably lighter meson mass. Since $M-\kappa>m$ in regard to the contributions from the finite Landau levels, as we used above, we have a weaker infrared expression $\tilde{I}_{1}=I_{1}/\l(M-\kappa\r)$ in the present work. 

Here $I_{2}$, which is related to the contributions from the finite landau levels, is canceled out by the valence quark mass $M$.
For simplicity, we rewrite the expression of 
\begin{align}
        \mb{B}= 2\mi g\tilde{J} \begin{pmatrix}\:
                1&1-\al\\
                1-\al&1-\be
        \:\end{pmatrix},
\end{align}
where $\tilde{J}=\l<J\r>_{0}+\l<K_{11}\r>_{n}$. $\al,\be$ are functions of $\l<J\r>_{0}$ and $\l<K_{(12,22)}\r>_{n}$ and $\al,\be\ll1$ if the magnitude of $\l<J\r>_{0}$ from the LLL is much larger than $\l<K_{(11,12,22)}\r>_{n}$ for $n\geq1$.

We obtain the roots of two pionic modes, $\hat{\pi}_{0}$ and $\bar{\pi}_{0}$, in the approximation of $(\eta,\zeta,\al,\be)\ll1$. Their perturbed masses take the forms of
\begin{align}
        m_{\hat{\pi}}^2&=\frac{1}{-2\mi g\tilde{J}}\frac{m}{M}
        +\frac{m+\kappa +\mi gI_{1}}{M}\frac{I_{1}}{M\tilde{J}}+\mc{O}(\al,\be);
\label{eqn:lightpion}\\
        m_{\bar{\pi}}^2&=\frac{1}{-\mi g\l(2\al-\be\r)\tilde{J}}+\mc{O}(\al^{0},\be^{0}).\label{eqn:heavypion}
\end{align}
Here $\mi I_{1}$ and $\mi\tilde{J}$ are positive and negative definite, respectively, since they originate from the loop integrations with  
one- and two-quark propagators.  
For the lighter pionic mode $\hat{\pi}$, its leading structure $m_{\hat{\pi}}^2=m/\l(-2\mi g\tilde{J}M\r)$ clearly reflects Goldstone’s theorem, as found earlier~\cite{Gusynin:1995nb,Avancini:2015ady}.
According to the second term of the r.h.s.~of \Eqn{eqn:lightpion}, the spectra of $\hat{\pi}$ is dramatically lowered by $\kappa$, which is in accordance with the numerical calculations in the work of~\cite{Xu:2020yag}.
It is interesting to remark that such mode behavior raises an interesting possibility of Bose condensation when the critical value
\begin{align}
        \kappa_{\mr{cr}}\simeq\frac{mM}{2\mi gI_{1}}-\mi gI_{1}-m,
\end{align}
is reached. {It is well known that in pion superfluidity the critical isospin chemical potential is equal to the pion mass at zero temperature and chemical potential~\cite{He:2005nk}. Therefore, in the limit $\mu_{I}^{c} = 0^{+}$, where the pion mass disappears due to massless quarks, the system immediately enters a Bose-Einstein Condensate (BEC) pion state. Similarly, regardless of how small $\kappa$ is, the conventional hadronic gas state becomes unstable and a new phase emerges in the chiral limit at $eB \neq 0$. Since several possible states have been proposed for strongly magnetized QCD matter~\cite{Ferrer:2015iop,Brauner:2016pko}, a sophisticated investigation of the vacuum state with more model parameters will be explored elsewhere.}

Due to the finite $\kappa$, the neutral pion fails to manifest itself as the Nambu-Goldstone boson. Such special nature is broken by a two-fold aspect. Firstly, the zero- and two-form pions couple to each other through the pseudo scalar-tensor bubbles, induced by the additional Dirac structures in the modified fermion propagator. Secondly, a newly developed infrared dynamics forms, as \Eqn{eqn:tildeI1} in the meson kernels. As $\kappa\to M$, the related LLL term is strongly enhanced, which reveals that the dimensional reduction is not restricted to the quark condensates, but is also present in the motion of the neutral excitation. As a remnant of the infrared dynamics, it is reasonable to recognize that a very small $\kappa$ is sufficient to reduce the mass of $\pi_{0}$ to zero. Of course, presented by the lattice simulations~\cite{Bali:2017ian,Bali:2018sey,Ding:2020jui,Ding:2020hxw}, the Bose condensation of neutral pion does not occur until the ultra-strong limit is reached, where $eB\sim 3.5\,\GeV^2$. Other additional factors are under exploration for a stronger strength of the anomalous coupling in a magnetized environment.
\section{Conclusion}\label{sec:con}
In the present work, we have employed the two-flavor NJL model to examine the properties of the generalized neutral pion in a constant background magnetic field.
Taking into account the back-reaction from the gluon sector, it allows us to decouple the longitudinal and transverse space and introduce the extra tensor-like four-fermion interactions in the model construction.
The novelty of the employed framework lies in that there are two order parameters emerging in the vacuum, described by $\l<\bar{\psi}\psi\r>\sim M$ and $\l<\bar{\psi}\si_{12}\psi\r>\sim \kappa$. Here $\kappa$ plays a role similar to the anomalous magnetic moment in the quark Dirac equation. 
As a spin dependent coupling, it is no wonder that $\kappa$ is not degenerate under the operation of charge conjugation like mass is. We prove its allowed sign is opposite to the sign of the quark's electric charge. 
Restricting to the model parameters where the coupling constants $G_{S}=G_{T}$,
we examine that the magnitude of $\kappa$ is smaller than the dynamical solution of the quark mass, as well.

Secondly, we revisit the qualitative description of the neutral pion under the influence of the AMM coupling. The key observation is that the ordinary meson is no longer the collective excitation of the system with multiple order parameters. A simultaneous treatment of fluctuations has to be implemented to realize the degrees of freedom of the meson modes. To the best of our knowledge, such a generalized neutral pion has not been examined in strong magnetic fields before. Properly including the pseudo-tensor vertex in RPA loop calculations, the spectra of the two pionic meson modes are presented after diagonalization. It is found that the familiar Goldstone nature is corrupted for the lightest chiral meson $\hat{\pi}_{0}$. Moreover, $\kappa$ strongly reduces its mass on two sides. On the one hand, the existence of the mixing always lowers one eigenvalue, but enhances another massive mode. On the other hand, we observe that a unique infrared dynamics arises in the meson correlator, which is cannot be labeled as the catalyzed dynamical quark mass. Hence, we point out that the treatment of the infrared cutoff will be very sensitive in the case of the nonrenormalized model calculations.
{Under a simple assumption, where the reduction from $D \to D-2$ affects only the charged channels, it is implied that the neutral $\pi_{0}$ is free to move in the original $3+1$ dimension and acts as an NG boson~\cite{Gusynin:1995nb}. However, as the system behaves like a $1+1$ dimension described by both one- and two-point correlators, it is likely that an inhomogeneous phase emerges, such as a chiral density wave state~\cite{Ferrer:2015iop} or a chiral soliton lattice~\cite{Brauner:2016pko}. A full numerical simulation is required to determine the phase state under the dimension reduction effect of the AMM coupling and will be calculated in the future.}

According to our results, the generalized pion continuously becomes lighter while $\kappa$ increases. Eventually, we expect that the interesting BEC of the generalized pion occurs when $\kappa$ is strong enough. Such exotic phase may be realized in a more complicated magnetized system~\cite{Chao:2018ejd,Vovchenko:2020crk}. We will discuss such a possibility in a future publication.

Rich phenomena have been reported in the present QCD$\times$QED environment. For example, the debate on the superconducting QCD vacuum~\cite{Chernodub:2011mc,Hidaka:2012mz}, the puzzle of magnetic susceptibility~\cite{Frasca:2011zn,Hofmann:2021bac,Xu:2020yag}, the understanding of the role of the pion mass in first-principles' simulations~\cite{DElia:2018xwo,Endrodi:2019zrl} and the strange metal phase of QCD in $1+1$ dimensions~\cite{Lajer:2021kcz}. 
Our approach including the effect of the AMM coupling has a potential to shed light on these discussions. We leave these projects to future works.
\begin{acknowledgements}
We thank S.B.Gudnason for his helpful feedback on the manuscript. JC was supported by the start-up funding from Jiangxi Normal University under Grant No. 12021211, while YXL was supported by the National Natural Science Foundation of China (NSFC) under Grant Nos. 12175007 and 12247107.
\end{acknowledgements}
\appendix
\section{Appendix: Technical Details of the Meson Kernel Matrix Element Calculation}

Considering the allowed kinematic regions of the pole, the first line (\ref{eqn:line-1}) in the bracket simplifies to
\begin{align}
(\ref{eqn:line-1})&=\l(1-\frac{p_{0}q_{0}-p_{3}q_{3}}{|p_{\sp}q_{\sp}|}\r)\frac{1}{\l(|p_{\sp}|-M+\kappa\r)\l(|q_{\sp}|-M+\kappa\r)}\nn
&=\frac{-\l(|p_{\sp}|-|q_{\sp}|\r)^2+m_{\pi}^{2}}{2|p_{\sp}q_{\sp}|}\frac{1}{|p_{\sp}|-|q_{\sp}|}
\l(\frac{1}{|q_{\sp}|-M+\kappa}-\frac{1}{|p_{\sp}|-M+\kappa}\r).
\end{align}
Similarly, one has
\begin{align}
(\ref{eqn:line-3})&=\sum_{\pm}\l(1+\frac{p_{0}q_{0}-p_{3}q_{3}}{|p_{\sp}q_{\sp}|}\r)\frac{-1}{\l(|p_{\sp}|\mp(M-\kappa)\r)\l(|q_{\sp}|\pm(M-\kappa)\r)}\nn
&=\frac{\l(|p_{\sp}|+|q_{\sp}|\r)^2-m_{\pi}^{2}}{2|p_{\sp}q_{\sp}|}
\frac{-1}{|p_{\sp}|+|q_{\sp}|}
\l(\frac{1}{|p_{\sp}|\mp(M-\kappa)}+\frac{1}{|q_{\sp}|\pm(M-\kappa)}\r).
\end{align}
To sum the two terms together, one has
\begin{align}\label{eqn:LLL}
        (\ref{eqn:line-1})+(\ref{eqn:line-3})&=\frac{-1}{|p_{\sp}|\l(|p_{\sp}|-M+\kappa\r)}+\frac{-1}{|q_{\sp}|\l(|q_{\sp}|-M+\kappa\r)}+\frac{m_{\pi}^{2}}{2|p_{\sp}q_{\sp}|}J
\end{align}
where
\begin{align}
        J=\frac{4|p_{\sp}q_{\sp}|}{\l[p_{\sp}^2-(M-\kappa)^2\r]\l[q_{\sp}^2-(M-\kappa)^2\r]}.
\end{align}

Stepping to the terms of finite Landau levels, it contains
\begin{align}
(\ref{eqn:line-5})
        &=\sum_{\pm}\sum_{n=1}^{\infty}\l(1-\frac{p_{0}q_{0}-p_{3}q_{3}}{|p_{\sp}q_{\sp}|}\r)
        \l(\frac{-1}{\l(|p_{\sp}|\pm\kappa\r)^{2}-M_{n}^{2}}+\frac{-1}{\l(|q_{\sp}|\pm\kappa\r)^{2}-M_{n}^{2}}\r)
        \nn
        &+\sum_{\pm}\sum_{n=1}^{\infty}\frac{m_{\pi}^{2}}{2|p_{\sp}q_{\sp}|}\l(1-\frac{4k_{0}^{2}}{\l(|p_{\sp}|+|q_{\sp}|\r)^2}\r)
        \frac{\l(|p_{\sp}|+|q_{\sp}|\pm2\kappa\r)^2}{\l[\l(|p_{\sp}|\pm\kappa\r)^{2}-M_{n}^{2}\r]\l[\l(|q_{\sp}|\pm\kappa\r)^{2}-M_{n}^{2}\r]},
\end{align}
and from the last line of
\begin{align}
(\ref{eqn:line-7})
        &=\sum_{\pm}\sum_{n=1}^{\infty}\l(1+\frac{p_{0}q_{0}-p_{3}q_{3}}{|p_{\sp}q_{\sp}|}\r)
        \l(\frac{-1}{\l(|p_{\sp}|\pm\kappa\r)^{2}-M_{n}^{2}}+\frac{-1}{\l(|q_{\sp}|\mp\kappa\r)^{2}-M_{n}^{2}}\r)
        \nn
        &+\sum_{\pm}\sum_{n=1}^{\infty}\frac{m_{\pi}^{2}}{2|p_{\sp}q_{\sp}|}\l(\frac{4k_{0}^{2}}{\l(|p_{\sp}|-|q_{\sp}|\r)^2}-1\r)
        \frac{\l(|p_{\sp}|-|q_{\sp}|\pm2\kappa\r)^2}{\l[\l(|p_{\sp}|\pm\kappa\r)^{2}-M_{n}^{2}\r]\l[\l(|q_{\sp}|\mp\kappa\r)^{2}-M_{n}^{2}\r]}.
\end{align}
Combining them together, one obtains that
\begin{align}
(\ref{eqn:line-5})+(\ref{eqn:line-7})&=\sum_{\pm}\sum_{n=1}^{\infty}\l[\frac{-2}{\l(|p_{\sp}|\pm\kappa\r)^{2}-M_{n}^{2}}+\frac{-2}{\l(|q_{\sp}|\pm\kappa\r)^{2}-M_{n}^{2}}+\frac{m_{\pi}^{2}}{2|p_{\sp}q_{\sp}|}K_{11}\r],
\end{align}
where 
\begin{align}
        K_{11}=&\l(1-\frac{4k_{0}^{2}}{\l(|p_{\sp}|+|q_{\sp}|\r)^2}\r)
        \frac{\l(|p_{\sp}|+|q_{\sp}|\pm2\kappa\r)^2}{\l[\l(|p_{\sp}|\pm\kappa\r)^{2}-M_{n}^{2}\r]\l[\l(|q_{\sp}|\pm\kappa\r)^{2}-M_{n}^{2}\r]}
\nn 
        &+\l(\frac{4k_{0}^{2}}{\l(|p_{\sp}|-|q_{\sp}|\r)^2}-1\r)
        \frac{\l(|p_{\sp}|-|q_{\sp}|\pm2\kappa\r)^2}{\l[\l(|p_{\sp}|\pm\kappa\r)^{2}-M_{n}^{2}\r]\l[\l(|q_{\sp}|\mp\kappa\r)^{2}-M_{n}^{2}\r]}.
\end{align}

Tracing in Dirac space, it is easy to get the mixed meson-meson correlators in the mixture of $\mi\ga_{5}\si_{12}\otimes\mi\ga_{5}\si_{12}$ and $\mi\ga_{5}\si_{12}\otimes\mi\ga_{5}$, described in terms of
\begin{align}
        K_{22}=&\l(1-\frac{4k_{0}^{2}}{\l(|p_{\sp}|+|q_{\sp}|\r)^2}\r)
        \frac{4M^2-\l(|p_{\sp}|-|q_{\sp}|\r)^2}
        {\l[\l(|p_{\sp}|\pm\kappa\r)^{2}-M_{n}^{2}\r]\l[\l(|q_{\sp}|\pm\kappa\r)^{2}-M_{n}^{2}\r]}\nn &
        +\l(\frac{4k_{0}^{2}}{\l(|p_{\sp}|-|q_{\sp}|\r)^2}-1\r)
        \frac{4M^2-\l(|p_{\sp}|+|q_{\sp}|\r)^2}
        {\l[\l(|p_{\sp}|\pm\kappa\r)^{2}-M_{n}^{2}\r]\l[\l(|q_{\sp}|\mp\kappa\r)^{2}-M_{n}^{2}\r]},
\end{align}
and
\begin{align}
        K_{12}=&\l(1-\frac{4k_{0}^{2}}{\l(|p_{\sp}|+|q_{\sp}|\r)^2}\r)
        \frac{\pm 2M\l(|p_{\sp}|+|q_{\sp}|\pm2\kappa\r)}
        {\l[\l(|p_{\sp}|\pm\kappa\r)^{2}-M_{n}^{2}\r]\l[\l(|q_{\sp}|\pm\kappa\r)^{2}-M_{n}^{2}\r]}\nn &
        +\l(\frac{4k_{0}^{2}}{\l(|p_{\sp}|-|q_{\sp}|\r)^2}-1\r)
        \frac{\pm 2M\l(|p_{\sp}|-|q_{\sp}|\pm2\kappa\r)}
        {\l[\l(|p_{\sp}|\pm\kappa\r)^{2}-M_{n}^{2}\r]\l[\l(|q_{\sp}|\mp\kappa\r)^{2}-M_{n}^{2}\r]}.
\end{align}

We note here that the term $\l(|p_{\sp}|-|q_{\sp}|\r)^2$ in the bracket will not lead to a new discussion of regularization for mesons, since its poles are located at i) $k_{0}=0$ for any $m_{\pi}$, which is not contributing due to the $k_{0}^2$ in the nominator; ii) $m_{\pi}=0$ for any $k_{0}$. In the latter case, the massless property of the neutral pion is guaranteed by the chiral quark and zero $\kappa$, hence, the explicit value of $K$ is not important at all.
%


\begin{thebibliography}{61}%
\makeatletter
\providecommand \@ifxundefined [1]{%
 \@ifx{#1\undefined}
}%
\providecommand \@ifnum [1]{%
 \ifnum #1\expandafter \@firstoftwo
 \else \expandafter \@secondoftwo
 \fi
}%
\providecommand \@ifx [1]{%
 \ifx #1\expandafter \@firstoftwo
 \else \expandafter \@secondoftwo
 \fi
}%
\providecommand \natexlab [1]{#1}%
\providecommand \enquote  [1]{``#1''}%
\providecommand \bibnamefont  [1]{#1}%
\providecommand \bibfnamefont [1]{#1}%
\providecommand \citenamefont [1]{#1}%
\providecommand \href@noop [0]{\@secondoftwo}%
\providecommand \href [0]{\begingroup \@sanitize@url \@href}%
\providecommand \@href[1]{\@@startlink{#1}\@@href}%
\providecommand \@@href[1]{\endgroup#1\@@endlink}%
\providecommand \@sanitize@url [0]{\catcode `\\12\catcode `\$12\catcode
  `\&12\catcode `\#12\catcode `\^12\catcode `\_12\catcode `\%12\relax}%
\providecommand \@@startlink[1]{}%
\providecommand \@@endlink[0]{}%
\providecommand \url  [0]{\begingroup\@sanitize@url \@url }%
\providecommand \@url [1]{\endgroup\@href {#1}{\urlprefix }}%
\providecommand \urlprefix  [0]{URL }%
\providecommand \Eprint [0]{\href }%
\providecommand \doibase [0]{https://doi.org/}%
\providecommand \selectlanguage [0]{\@gobble}%
\providecommand \bibinfo  [0]{\@secondoftwo}%
\providecommand \bibfield  [0]{\@secondoftwo}%
\providecommand \translation [1]{[#1]}%
\providecommand \BibitemOpen [0]{}%
\providecommand \bibitemStop [0]{}%
\providecommand \bibitemNoStop [0]{.\EOS\space}%
\providecommand \EOS [0]{\spacefactor3000\relax}%
\providecommand \BibitemShut  [1]{\csname bibitem#1\endcsname}%
\let\auto@bib@innerbib\@empty
\bibitem [{\citenamefont {Bali}\ \emph
  {et~al.}(2018{\natexlab{a}})\citenamefont {Bali}, \citenamefont {Brandt},
  \citenamefont {Endr\H{o}di},\ and\ \citenamefont
  {Gl\"a\ss{}le}}]{Bali:2018sey}%
  \BibitemOpen
  \bibfield  {author} {\bibinfo {author} {\bibfnamefont {G.~S.}\ \bibnamefont
  {Bali}}, \bibinfo {author} {\bibfnamefont {B.~B.}\ \bibnamefont {Brandt}},
  \bibinfo {author} {\bibfnamefont {G.}~\bibnamefont {Endr\H{o}di}},\ and\
  \bibinfo {author} {\bibfnamefont {B.}~\bibnamefont {Gl\"a\ss{}le}},\
  }\bibfield  {title} {\bibinfo {title} {{Weak decay of magnetized pions}},\
  }\href {https://doi.org/10.1103/PhysRevLett.121.072001} {\bibfield  {journal}
  {\bibinfo  {journal} {Phys. Rev. Lett.}\ }\textbf {\bibinfo {volume} {121}},\
  \bibinfo {pages} {072001} (\bibinfo {year} {2018}{\natexlab{a}})},\ \Eprint
  {https://arxiv.org/abs/1805.10971} {arXiv:1805.10971 [hep-lat]} \BibitemShut
  {NoStop}%
\bibitem [{\citenamefont {Bali}\ \emph
  {et~al.}(2018{\natexlab{b}})\citenamefont {Bali}, \citenamefont {Brandt},
  \citenamefont {Endr\H{o}di},\ and\ \citenamefont
  {Gl\"a\ss{}le}}]{Bali:2017ian}%
  \BibitemOpen
  \bibfield  {author} {\bibinfo {author} {\bibfnamefont {G.~S.}\ \bibnamefont
  {Bali}}, \bibinfo {author} {\bibfnamefont {B.~B.}\ \bibnamefont {Brandt}},
  \bibinfo {author} {\bibfnamefont {G.}~\bibnamefont {Endr\H{o}di}},\ and\
  \bibinfo {author} {\bibfnamefont {B.}~\bibnamefont {Gl\"a\ss{}le}},\
  }\bibfield  {title} {\bibinfo {title} {{Meson masses in electromagnetic
  fields with Wilson fermions}},\ }\href
  {https://doi.org/10.1103/PhysRevD.97.034505} {\bibfield  {journal} {\bibinfo
  {journal} {Phys. Rev. D}\ }\textbf {\bibinfo {volume} {97}},\ \bibinfo
  {pages} {034505} (\bibinfo {year} {2018}{\natexlab{b}})},\ \Eprint
  {https://arxiv.org/abs/1707.05600} {arXiv:1707.05600 [hep-lat]} \BibitemShut
  {NoStop}%
\bibitem [{\citenamefont {Ding}\ \emph {et~al.}(2020)\citenamefont {Ding},
  \citenamefont {Li}, \citenamefont {Mukherjee}, \citenamefont {Tomiya},\ and\
  \citenamefont {Wang}}]{Ding:2020jui}%
  \BibitemOpen
  \bibfield  {author} {\bibinfo {author} {\bibfnamefont {H.-T.}\ \bibnamefont
  {Ding}}, \bibinfo {author} {\bibfnamefont {S.-T.}\ \bibnamefont {Li}},
  \bibinfo {author} {\bibfnamefont {S.}~\bibnamefont {Mukherjee}}, \bibinfo
  {author} {\bibfnamefont {A.}~\bibnamefont {Tomiya}},\ and\ \bibinfo {author}
  {\bibfnamefont {X.-D.}\ \bibnamefont {Wang}},\ }\bibfield  {title} {\bibinfo
  {title} {{Meson masses in external magnetic fields with HISQ fermions}},\
  }\href {https://doi.org/10.22323/1.363.0250} {\bibfield  {journal} {\bibinfo
  {journal} {PoS}\ }\textbf {\bibinfo {volume} {LATTICE2019}},\ \bibinfo
  {pages} {250} (\bibinfo {year} {2020})},\ \Eprint
  {https://arxiv.org/abs/2001.05322} {arXiv:2001.05322 [hep-lat]} \BibitemShut
  {NoStop}%
\bibitem [{\citenamefont {Ding}\ \emph {et~al.}(2021)\citenamefont {Ding},
  \citenamefont {Li}, \citenamefont {Tomiya}, \citenamefont {Wang},\ and\
  \citenamefont {Zhang}}]{Ding:2020hxw}%
  \BibitemOpen
  \bibfield  {author} {\bibinfo {author} {\bibfnamefont {H.~T.}\ \bibnamefont
  {Ding}}, \bibinfo {author} {\bibfnamefont {S.~T.}\ \bibnamefont {Li}},
  \bibinfo {author} {\bibfnamefont {A.}~\bibnamefont {Tomiya}}, \bibinfo
  {author} {\bibfnamefont {X.~D.}\ \bibnamefont {Wang}},\ and\ \bibinfo
  {author} {\bibfnamefont {Y.}~\bibnamefont {Zhang}},\ }\bibfield  {title}
  {\bibinfo {title} {{Chiral properties of (2+1)-flavor QCD in strong magnetic
  fields at zero temperature}},\ }\href
  {https://doi.org/10.1103/PhysRevD.104.014505} {\bibfield  {journal} {\bibinfo
   {journal} {Phys. Rev. D}\ }\textbf {\bibinfo {volume} {104}},\ \bibinfo
  {pages} {014505} (\bibinfo {year} {2021})},\ \Eprint
  {https://arxiv.org/abs/2008.00493} {arXiv:2008.00493 [hep-lat]} \BibitemShut
  {NoStop}%
\bibitem [{\citenamefont {Skokov}\ \emph {et~al.}(2009)\citenamefont {Skokov},
  \citenamefont {Illarionov},\ and\ \citenamefont {Toneev}}]{Skokov:2009qp}%
  \BibitemOpen
  \bibfield  {author} {\bibinfo {author} {\bibfnamefont {V.}~\bibnamefont
  {Skokov}}, \bibinfo {author} {\bibfnamefont {A.~Y.}\ \bibnamefont
  {Illarionov}},\ and\ \bibinfo {author} {\bibfnamefont {V.}~\bibnamefont
  {Toneev}},\ }\bibfield  {title} {\bibinfo {title} {{Estimate of the magnetic
  field strength in heavy-ion collisions}},\ }\href
  {https://doi.org/10.1142/S0217751X09047570} {\bibfield  {journal} {\bibinfo
  {journal} {Int. J. Mod. Phys. A}\ }\textbf {\bibinfo {volume} {24}},\
  \bibinfo {pages} {5925} (\bibinfo {year} {2009})},\ \Eprint
  {https://arxiv.org/abs/0907.1396} {arXiv:0907.1396 [nucl-th]} \BibitemShut
  {NoStop}%
\bibitem [{\citenamefont {Deng}\ and\ \citenamefont
  {Huang}(2012)}]{Deng:2012pc}%
  \BibitemOpen
  \bibfield  {author} {\bibinfo {author} {\bibfnamefont {W.-T.}\ \bibnamefont
  {Deng}}\ and\ \bibinfo {author} {\bibfnamefont {X.-G.}\ \bibnamefont
  {Huang}},\ }\bibfield  {title} {\bibinfo {title} {{Event-by-event generation
  of electromagnetic fields in heavy-ion collisions}},\ }\href
  {https://doi.org/10.1103/PhysRevC.85.044907} {\bibfield  {journal} {\bibinfo
  {journal} {Phys. Rev. C}\ }\textbf {\bibinfo {volume} {85}},\ \bibinfo
  {pages} {044907} (\bibinfo {year} {2012})},\ \Eprint
  {https://arxiv.org/abs/1201.5108} {arXiv:1201.5108 [nucl-th]} \BibitemShut
  {NoStop}%
\bibitem [{\citenamefont {Kharzeev}\ \emph {et~al.}(2013)\citenamefont
  {Kharzeev}, \citenamefont {Landsteiner}, \citenamefont {Schmitt},\ and\
  \citenamefont {Yee}}]{Kharzeev:2012ph}%
  \BibitemOpen
  \bibfield  {author} {\bibinfo {author} {\bibfnamefont {D.~E.}\ \bibnamefont
  {Kharzeev}}, \bibinfo {author} {\bibfnamefont {K.}~\bibnamefont
  {Landsteiner}}, \bibinfo {author} {\bibfnamefont {A.}~\bibnamefont
  {Schmitt}},\ and\ \bibinfo {author} {\bibfnamefont {H.-U.}\ \bibnamefont
  {Yee}},\ }\bibfield  {title} {\bibinfo {title} {{'Strongly interacting matter
  in magnetic fields': an overview}},\ }\href
  {https://doi.org/10.1007/978-3-642-37305-3_1} {\bibfield  {journal} {\bibinfo
   {journal} {Lect. Notes Phys.}\ }\textbf {\bibinfo {volume} {871}},\ \bibinfo
  {pages} {1} (\bibinfo {year} {2013})},\ \Eprint
  {https://arxiv.org/abs/1211.6245} {arXiv:1211.6245 [hep-ph]} \BibitemShut
  {NoStop}%
\bibitem [{\citenamefont {Kaspi}\ and\ \citenamefont
  {Beloborodov}(2017)}]{Kaspi:2017fwg}%
  \BibitemOpen
  \bibfield  {author} {\bibinfo {author} {\bibfnamefont {V.~M.}\ \bibnamefont
  {Kaspi}}\ and\ \bibinfo {author} {\bibfnamefont {A.}~\bibnamefont
  {Beloborodov}},\ }\bibfield  {title} {\bibinfo {title} {{Magnetars}},\ }\href
  {https://doi.org/10.1146/annurev-astro-081915-023329} {\bibfield  {journal}
  {\bibinfo  {journal} {Ann. Rev. Astron. Astrophys.}\ }\textbf {\bibinfo
  {volume} {55}},\ \bibinfo {pages} {261} (\bibinfo {year} {2017})},\ \Eprint
  {https://arxiv.org/abs/1703.00068} {arXiv:1703.00068 [astro-ph.HE]}
  \BibitemShut {NoStop}%
\bibitem [{\citenamefont {Andersen}(2012{\natexlab{a}})}]{Andersen:2012zc}%
  \BibitemOpen
  \bibfield  {author} {\bibinfo {author} {\bibfnamefont {J.~O.}\ \bibnamefont
  {Andersen}},\ }\bibfield  {title} {\bibinfo {title} {{Chiral perturbation
  theory in a magnetic background - finite-temperature effects}},\ }\href
  {https://doi.org/10.1007/JHEP10(2012)005} {\bibfield  {journal} {\bibinfo
  {journal} {JHEP}\ }\textbf {\bibinfo {volume} {10}},\ \bibinfo {pages}
  {005}},\ \Eprint {https://arxiv.org/abs/1205.6978} {arXiv:1205.6978 [hep-ph]}
  \BibitemShut {NoStop}%
\bibitem [{\citenamefont {Chao}(2020)}]{Chao:2018czo}%
  \BibitemOpen
  \bibfield  {author} {\bibinfo {author} {\bibfnamefont {J.}~\bibnamefont
  {Chao}},\ }\bibfield  {title} {\bibinfo {title} {{Phase diagram of two-color
  QCD matter at finite baryon and axial isospin densities}},\ }\href
  {https://doi.org/10.1088/1674-1137/44/3/034108} {\bibfield  {journal}
  {\bibinfo  {journal} {Chin. Phys. C}\ }\textbf {\bibinfo {volume} {44}},\
  \bibinfo {pages} {034108} (\bibinfo {year} {2020})},\ \Eprint
  {https://arxiv.org/abs/1808.01928} {arXiv:1808.01928 [hep-ph]} \BibitemShut
  {NoStop}%
\bibitem [{\citenamefont {Andersen}(2012{\natexlab{b}})}]{Andersen:2012dz}%
  \BibitemOpen
  \bibfield  {author} {\bibinfo {author} {\bibfnamefont {J.~O.}\ \bibnamefont
  {Andersen}},\ }\bibfield  {title} {\bibinfo {title} {{Thermal pions in a
  magnetic background}},\ }\href {https://doi.org/10.1103/PhysRevD.86.025020}
  {\bibfield  {journal} {\bibinfo  {journal} {Phys. Rev. D}\ }\textbf {\bibinfo
  {volume} {86}},\ \bibinfo {pages} {025020} (\bibinfo {year}
  {2012}{\natexlab{b}})},\ \Eprint {https://arxiv.org/abs/1202.2051}
  {arXiv:1202.2051 [hep-ph]} \BibitemShut {NoStop}%
\bibitem [{\citenamefont {Gorsky}\ \emph {et~al.}(2012)\citenamefont {Gorsky},
  \citenamefont {Kopnin}, \citenamefont {Krikun},\ and\ \citenamefont
  {Vainshtein}}]{Gorsky:2012ui}%
  \BibitemOpen
  \bibfield  {author} {\bibinfo {author} {\bibfnamefont {A.}~\bibnamefont
  {Gorsky}}, \bibinfo {author} {\bibfnamefont {P.~N.}\ \bibnamefont {Kopnin}},
  \bibinfo {author} {\bibfnamefont {A.}~\bibnamefont {Krikun}},\ and\ \bibinfo
  {author} {\bibfnamefont {A.}~\bibnamefont {Vainshtein}},\ }\bibfield  {title}
  {\bibinfo {title} {{More on the Tensor Response of the QCD Vacuum to an
  External Magnetic Field}},\ }\href
  {https://doi.org/10.1103/PhysRevD.85.086006} {\bibfield  {journal} {\bibinfo
  {journal} {Phys. Rev. D}\ }\textbf {\bibinfo {volume} {85}},\ \bibinfo
  {pages} {086006} (\bibinfo {year} {2012})},\ \Eprint
  {https://arxiv.org/abs/1201.2039} {arXiv:1201.2039 [hep-ph]} \BibitemShut
  {NoStop}%
\bibitem [{\citenamefont {Avancini}\ \emph {et~al.}(2017)\citenamefont
  {Avancini}, \citenamefont {Farias}, \citenamefont {Benghi~Pinto},
  \citenamefont {Tavares},\ and\ \citenamefont {Tim\'oteo}}]{Avancini:2016fgq}%
  \BibitemOpen
  \bibfield  {author} {\bibinfo {author} {\bibfnamefont {S.~S.}\ \bibnamefont
  {Avancini}}, \bibinfo {author} {\bibfnamefont {R.~L.~S.}\ \bibnamefont
  {Farias}}, \bibinfo {author} {\bibfnamefont {M.}~\bibnamefont
  {Benghi~Pinto}}, \bibinfo {author} {\bibfnamefont {W.~R.}\ \bibnamefont
  {Tavares}},\ and\ \bibinfo {author} {\bibfnamefont {V.~S.}\ \bibnamefont
  {Tim\'oteo}},\ }\bibfield  {title} {\bibinfo {title} {{$\pi_0$ pole mass
  calculation in a strong magnetic field and lattice constraints}},\ }\href
  {https://doi.org/10.1016/j.physletb.2017.02.002} {\bibfield  {journal}
  {\bibinfo  {journal} {Phys. Lett. B}\ }\textbf {\bibinfo {volume} {767}},\
  \bibinfo {pages} {247} (\bibinfo {year} {2017})},\ \Eprint
  {https://arxiv.org/abs/1606.05754} {arXiv:1606.05754 [hep-ph]} \BibitemShut
  {NoStop}%
\bibitem [{\citenamefont {Zhang}\ \emph {et~al.}(2016)\citenamefont {Zhang},
  \citenamefont {Fu},\ and\ \citenamefont {Liu}}]{Zhang:2016qrl}%
  \BibitemOpen
  \bibfield  {author} {\bibinfo {author} {\bibfnamefont {R.}~\bibnamefont
  {Zhang}}, \bibinfo {author} {\bibfnamefont {W.-j.}\ \bibnamefont {Fu}},\ and\
  \bibinfo {author} {\bibfnamefont {Y.-x.}\ \bibnamefont {Liu}},\ }\bibfield
  {title} {\bibinfo {title} {{Properties of Mesons in a Strong Magnetic
  Field}},\ }\href {https://doi.org/10.1140/epjc/s10052-016-4123-8} {\bibfield
  {journal} {\bibinfo  {journal} {Eur. Phys. J. C}\ }\textbf {\bibinfo {volume}
  {76}},\ \bibinfo {pages} {307} (\bibinfo {year} {2016})},\ \Eprint
  {https://arxiv.org/abs/1604.08888} {arXiv:1604.08888 [hep-ph]} \BibitemShut
  {NoStop}%
\bibitem [{\citenamefont {Wang}\ and\ \citenamefont
  {Zhuang}(2018)}]{Wang:2017vtn}%
  \BibitemOpen
  \bibfield  {author} {\bibinfo {author} {\bibfnamefont {Z.}~\bibnamefont
  {Wang}}\ and\ \bibinfo {author} {\bibfnamefont {P.}~\bibnamefont {Zhuang}},\
  }\bibfield  {title} {\bibinfo {title} {{Meson properties in magnetized quark
  matter}},\ }\href {https://doi.org/10.1103/PhysRevD.97.034026} {\bibfield
  {journal} {\bibinfo  {journal} {Phys. Rev. D}\ }\textbf {\bibinfo {volume}
  {97}},\ \bibinfo {pages} {034026} (\bibinfo {year} {2018})},\ \Eprint
  {https://arxiv.org/abs/1712.00554} {arXiv:1712.00554 [hep-ph]} \BibitemShut
  {NoStop}%
\bibitem [{\citenamefont {Liu}\ \emph {et~al.}(2018)\citenamefont {Liu},
  \citenamefont {Wang}, \citenamefont {Yu},\ and\ \citenamefont
  {Huang}}]{Liu:2018zag}%
  \BibitemOpen
  \bibfield  {author} {\bibinfo {author} {\bibfnamefont {H.}~\bibnamefont
  {Liu}}, \bibinfo {author} {\bibfnamefont {X.}~\bibnamefont {Wang}}, \bibinfo
  {author} {\bibfnamefont {L.}~\bibnamefont {Yu}},\ and\ \bibinfo {author}
  {\bibfnamefont {M.}~\bibnamefont {Huang}},\ }\bibfield  {title} {\bibinfo
  {title} {{Neutral and charged scalar mesons, pseudoscalar mesons, and
  diquarks in magnetic fields}},\ }\href
  {https://doi.org/10.1103/PhysRevD.97.076008} {\bibfield  {journal} {\bibinfo
  {journal} {Phys. Rev. D}\ }\textbf {\bibinfo {volume} {97}},\ \bibinfo
  {pages} {076008} (\bibinfo {year} {2018})},\ \Eprint
  {https://arxiv.org/abs/1801.02174} {arXiv:1801.02174 [hep-ph]} \BibitemShut
  {NoStop}%
\bibitem [{\citenamefont {Mao}(2019)}]{Mao:2018dqe}%
  \BibitemOpen
  \bibfield  {author} {\bibinfo {author} {\bibfnamefont {S.}~\bibnamefont
  {Mao}},\ }\bibfield  {title} {\bibinfo {title} {{Pions in magnetic field at
  finite temperature}},\ }\href {https://doi.org/10.1103/PhysRevD.99.056005}
  {\bibfield  {journal} {\bibinfo  {journal} {Phys. Rev. D}\ }\textbf {\bibinfo
  {volume} {99}},\ \bibinfo {pages} {056005} (\bibinfo {year} {2019})},\
  \Eprint {https://arxiv.org/abs/1808.10242} {arXiv:1808.10242 [nucl-th]}
  \BibitemShut {NoStop}%
\bibitem [{\citenamefont {Coppola}\ \emph {et~al.}(2019)\citenamefont
  {Coppola}, \citenamefont {Gomez~Dumm}, \citenamefont {Noguera},\ and\
  \citenamefont {Scoccola}}]{Coppola:2019uyr}%
  \BibitemOpen
  \bibfield  {author} {\bibinfo {author} {\bibfnamefont {M.}~\bibnamefont
  {Coppola}}, \bibinfo {author} {\bibfnamefont {D.}~\bibnamefont {Gomez~Dumm}},
  \bibinfo {author} {\bibfnamefont {S.}~\bibnamefont {Noguera}},\ and\ \bibinfo
  {author} {\bibfnamefont {N.~N.}\ \bibnamefont {Scoccola}},\ }\bibfield
  {title} {\bibinfo {title} {{Neutral and charged pion properties under strong
  magnetic fields in the NJL model}},\ }\href
  {https://doi.org/10.1103/PhysRevD.100.054014} {\bibfield  {journal} {\bibinfo
   {journal} {Phys. Rev. D}\ }\textbf {\bibinfo {volume} {100}},\ \bibinfo
  {pages} {054014} (\bibinfo {year} {2019})},\ \Eprint
  {https://arxiv.org/abs/1907.05840} {arXiv:1907.05840 [hep-ph]} \BibitemShut
  {NoStop}%
\bibitem [{\citenamefont {Das}\ and\ \citenamefont
  {Haque}(2020)}]{Das:2019ehv}%
  \BibitemOpen
  \bibfield  {author} {\bibinfo {author} {\bibfnamefont {A.}~\bibnamefont
  {Das}}\ and\ \bibinfo {author} {\bibfnamefont {N.}~\bibnamefont {Haque}},\
  }\bibfield  {title} {\bibinfo {title} {{Neutral pion mass in the linear sigma
  model coupled to quarks at arbitrary magnetic field}},\ }\href
  {https://doi.org/10.1103/PhysRevD.101.074033} {\bibfield  {journal} {\bibinfo
   {journal} {Phys. Rev. D}\ }\textbf {\bibinfo {volume} {101}},\ \bibinfo
  {pages} {074033} (\bibinfo {year} {2020})},\ \Eprint
  {https://arxiv.org/abs/1908.10323} {arXiv:1908.10323 [hep-ph]} \BibitemShut
  {NoStop}%
\bibitem [{\citenamefont {Sheng}\ \emph {et~al.}(2021)\citenamefont {Sheng},
  \citenamefont {Wang}, \citenamefont {Wang},\ and\ \citenamefont
  {Yu}}]{Sheng:2020hge}%
  \BibitemOpen
  \bibfield  {author} {\bibinfo {author} {\bibfnamefont {B.}~\bibnamefont
  {Sheng}}, \bibinfo {author} {\bibfnamefont {Y.}~\bibnamefont {Wang}},
  \bibinfo {author} {\bibfnamefont {X.}~\bibnamefont {Wang}},\ and\ \bibinfo
  {author} {\bibfnamefont {L.}~\bibnamefont {Yu}},\ }\bibfield  {title}
  {\bibinfo {title} {{Pole and screening masses of neutral pions in a hot and
  magnetized medium: A comprehensive study in the
  Nambu\textendash{}Jona-Lasinio model}},\ }\href
  {https://doi.org/10.1103/PhysRevD.103.094001} {\bibfield  {journal} {\bibinfo
   {journal} {Phys. Rev. D}\ }\textbf {\bibinfo {volume} {103}},\ \bibinfo
  {pages} {094001} (\bibinfo {year} {2021})},\ \Eprint
  {https://arxiv.org/abs/2010.05716} {arXiv:2010.05716 [hep-ph]} \BibitemShut
  {NoStop}%
\bibitem [{\citenamefont {Ayala}\ \emph {et~al.}(2021)\citenamefont {Ayala},
  \citenamefont {Hern\'andez}, \citenamefont {Hern\'andez}, \citenamefont
  {Farias},\ and\ \citenamefont {Zamora}}]{Ayala:2020dxs}%
  \BibitemOpen
  \bibfield  {author} {\bibinfo {author} {\bibfnamefont {A.}~\bibnamefont
  {Ayala}}, \bibinfo {author} {\bibfnamefont {J.~L.}\ \bibnamefont
  {Hern\'andez}}, \bibinfo {author} {\bibfnamefont {L.~A.}\ \bibnamefont
  {Hern\'andez}}, \bibinfo {author} {\bibfnamefont {R.~L.~S.}\ \bibnamefont
  {Farias}},\ and\ \bibinfo {author} {\bibfnamefont {R.}~\bibnamefont
  {Zamora}},\ }\bibfield  {title} {\bibinfo {title} {{Magnetic field dependence
  of the neutral pion mass in the linear sigma model with quarks: The strong
  field case}},\ }\href {https://doi.org/10.1103/PhysRevD.103.054038}
  {\bibfield  {journal} {\bibinfo  {journal} {Phys. Rev. D}\ }\textbf {\bibinfo
  {volume} {103}},\ \bibinfo {pages} {054038} (\bibinfo {year} {2021})},\
  \Eprint {https://arxiv.org/abs/2011.03673} {arXiv:2011.03673 [hep-ph]}
  \BibitemShut {NoStop}%
\bibitem [{\citenamefont {Xing}\ \emph {et~al.}(2022)\citenamefont {Xing},
  \citenamefont {Chao}, \citenamefont {Chang},\ and\ \citenamefont
  {Liu}}]{Xing:2021kbw}%
  \BibitemOpen
  \bibfield  {author} {\bibinfo {author} {\bibfnamefont {Z.}~\bibnamefont
  {Xing}}, \bibinfo {author} {\bibfnamefont {J.}~\bibnamefont {Chao}}, \bibinfo
  {author} {\bibfnamefont {L.}~\bibnamefont {Chang}},\ and\ \bibinfo {author}
  {\bibfnamefont {Y.-x.}\ \bibnamefont {Liu}},\ }\bibfield  {title} {\bibinfo
  {title} {{Exposing the effect of the p-wave component in the pion triplet
  under a strong magnetic field}},\ }\href
  {https://doi.org/10.1103/PhysRevD.105.114003} {\bibfield  {journal} {\bibinfo
   {journal} {Phys. Rev. D}\ }\textbf {\bibinfo {volume} {105}},\ \bibinfo
  {pages} {114003} (\bibinfo {year} {2022})},\ \Eprint
  {https://arxiv.org/abs/2110.01245} {arXiv:2110.01245 [hep-ph]} \BibitemShut
  {NoStop}%
\bibitem [{\citenamefont {Ferrer}\ and\ \citenamefont {de~la
  Incera}(2009)}]{Ferrer:2008dy}%
  \BibitemOpen
  \bibfield  {author} {\bibinfo {author} {\bibfnamefont {E.~J.}\ \bibnamefont
  {Ferrer}}\ and\ \bibinfo {author} {\bibfnamefont {V.}~\bibnamefont {de~la
  Incera}},\ }\bibfield  {title} {\bibinfo {title} {{Dynamically Induced Zeeman
  Effect in Massless QED}},\ }\href
  {https://doi.org/10.1103/PhysRevLett.102.050402} {\bibfield  {journal}
  {\bibinfo  {journal} {Phys. Rev. Lett.}\ }\textbf {\bibinfo {volume} {102}},\
  \bibinfo {pages} {050402} (\bibinfo {year} {2009})},\ \Eprint
  {https://arxiv.org/abs/0807.4744} {arXiv:0807.4744 [hep-ph]} \BibitemShut
  {NoStop}%
\bibitem [{\citenamefont {Bueno}\ \emph {et~al.}(2012)\citenamefont {Bueno},
  \citenamefont {Furtado},\ and\ \citenamefont {Carvalho}}]{bueno2012landau}%
  \BibitemOpen
  \bibfield  {author} {\bibinfo {author} {\bibfnamefont {M.}~\bibnamefont
  {Bueno}}, \bibinfo {author} {\bibfnamefont {C.}~\bibnamefont {Furtado}},\
  and\ \bibinfo {author} {\bibfnamefont {A.~d.~M.}\ \bibnamefont {Carvalho}},\
  }\bibfield  {title} {\bibinfo {title} {Landau levels in graphene layers with
  topological defects},\ }\href@noop {} {\bibfield  {journal} {\bibinfo
  {journal} {The European Physical Journal B}\ }\textbf {\bibinfo {volume}
  {85}},\ \bibinfo {pages} {1} (\bibinfo {year} {2012})}\BibitemShut {NoStop}%
\bibitem [{\citenamefont {Eminov}(2018)}]{Eminov:2018lse}%
  \BibitemOpen
  \bibfield  {author} {\bibinfo {author} {\bibfnamefont {P.~A.}\ \bibnamefont
  {Eminov}},\ }\bibfield  {title} {\bibinfo {title} {{Anomalous magnetic moment
  of an electron in a constant magnetic field in ( 2+1 )-dimensional quantum
  electrodynamics}},\ }\href {https://doi.org/10.1103/PhysRevD.97.095019}
  {\bibfield  {journal} {\bibinfo  {journal} {Phys. Rev. D}\ }\textbf {\bibinfo
  {volume} {97}},\ \bibinfo {pages} {095019} (\bibinfo {year}
  {2018})}\BibitemShut {NoStop}%
\bibitem [{\citenamefont {Bicudo}\ \emph {et~al.}(1999)\citenamefont {Bicudo},
  \citenamefont {Ribeiro},\ and\ \citenamefont {Fernandes}}]{Bicudo:1998qb}%
  \BibitemOpen
  \bibfield  {author} {\bibinfo {author} {\bibfnamefont {P.~J.~A.}\
  \bibnamefont {Bicudo}}, \bibinfo {author} {\bibfnamefont {J.~E. F.~T.}\
  \bibnamefont {Ribeiro}},\ and\ \bibinfo {author} {\bibfnamefont
  {R.}~\bibnamefont {Fernandes}},\ }\bibfield  {title} {\bibinfo {title} {{The
  Anomalous magnetic moment of quarks}},\ }\href
  {https://doi.org/10.1103/PhysRevC.59.1107} {\bibfield  {journal} {\bibinfo
  {journal} {Phys. Rev. C}\ }\textbf {\bibinfo {volume} {59}},\ \bibinfo
  {pages} {1107} (\bibinfo {year} {1999})},\ \Eprint
  {https://arxiv.org/abs/hep-ph/9806243} {arXiv:hep-ph/9806243} \BibitemShut
  {NoStop}%
\bibitem [{\citenamefont {Faccioli}\ \emph {et~al.}(2003)\citenamefont
  {Faccioli}, \citenamefont {Schwenk},\ and\ \citenamefont
  {Shuryak}}]{Faccioli:2002jd}%
  \BibitemOpen
  \bibfield  {author} {\bibinfo {author} {\bibfnamefont {P.}~\bibnamefont
  {Faccioli}}, \bibinfo {author} {\bibfnamefont {A.}~\bibnamefont {Schwenk}},\
  and\ \bibinfo {author} {\bibfnamefont {E.~V.}\ \bibnamefont {Shuryak}},\
  }\bibfield  {title} {\bibinfo {title} {{Instanton contribution to the pion
  electromagnetic form-factor at Q**2 greater than 1-GeV**2}},\ }\href
  {https://doi.org/10.1103/PhysRevD.67.113009} {\bibfield  {journal} {\bibinfo
  {journal} {Phys. Rev. D}\ }\textbf {\bibinfo {volume} {67}},\ \bibinfo
  {pages} {113009} (\bibinfo {year} {2003})},\ \Eprint
  {https://arxiv.org/abs/hep-ph/0202027} {arXiv:hep-ph/0202027} \BibitemShut
  {NoStop}%
\bibitem [{\citenamefont {Chang}\ \emph {et~al.}(2011)\citenamefont {Chang},
  \citenamefont {Liu},\ and\ \citenamefont {Roberts}}]{Chang:2010hb}%
  \BibitemOpen
  \bibfield  {author} {\bibinfo {author} {\bibfnamefont {L.}~\bibnamefont
  {Chang}}, \bibinfo {author} {\bibfnamefont {Y.-X.}\ \bibnamefont {Liu}},\
  and\ \bibinfo {author} {\bibfnamefont {C.~D.}\ \bibnamefont {Roberts}},\
  }\bibfield  {title} {\bibinfo {title} {{Dressed-quark anomalous magnetic
  moments}},\ }\href {https://doi.org/10.1103/PhysRevLett.106.072001}
  {\bibfield  {journal} {\bibinfo  {journal} {Phys. Rev. Lett.}\ }\textbf
  {\bibinfo {volume} {106}},\ \bibinfo {pages} {072001} (\bibinfo {year}
  {2011})},\ \Eprint {https://arxiv.org/abs/1009.3458} {arXiv:1009.3458
  [nucl-th]} \BibitemShut {NoStop}%
\bibitem [{\citenamefont {Chang}\ \emph {et~al.}(2013)\citenamefont {Chang},
  \citenamefont {Clo\"et}, \citenamefont {Roberts}, \citenamefont {Schmidt},\
  and\ \citenamefont {Tandy}}]{Chang:2013nia}%
  \BibitemOpen
  \bibfield  {author} {\bibinfo {author} {\bibfnamefont {L.}~\bibnamefont
  {Chang}}, \bibinfo {author} {\bibfnamefont {I.~C.}\ \bibnamefont {Clo\"et}},
  \bibinfo {author} {\bibfnamefont {C.~D.}\ \bibnamefont {Roberts}}, \bibinfo
  {author} {\bibfnamefont {S.~M.}\ \bibnamefont {Schmidt}},\ and\ \bibinfo
  {author} {\bibfnamefont {P.~C.}\ \bibnamefont {Tandy}},\ }\bibfield  {title}
  {\bibinfo {title} {{Pion electromagnetic form factor at spacelike momenta}},\
  }\href {https://doi.org/10.1103/PhysRevLett.111.141802} {\bibfield  {journal}
  {\bibinfo  {journal} {Phys. Rev. Lett.}\ }\textbf {\bibinfo {volume} {111}},\
  \bibinfo {pages} {141802} (\bibinfo {year} {2013})},\ \Eprint
  {https://arxiv.org/abs/1307.0026} {arXiv:1307.0026 [nucl-th]} \BibitemShut
  {NoStop}%
\bibitem [{\citenamefont {Gutsche}\ \emph {et~al.}(2015)\citenamefont
  {Gutsche}, \citenamefont {Lyubovitskij}, \citenamefont {Schmidt},\ and\
  \citenamefont {Vega}}]{Gutsche:2014zua}%
  \BibitemOpen
  \bibfield  {author} {\bibinfo {author} {\bibfnamefont {T.}~\bibnamefont
  {Gutsche}}, \bibinfo {author} {\bibfnamefont {V.~E.}\ \bibnamefont
  {Lyubovitskij}}, \bibinfo {author} {\bibfnamefont {I.}~\bibnamefont
  {Schmidt}},\ and\ \bibinfo {author} {\bibfnamefont {A.}~\bibnamefont
  {Vega}},\ }\bibfield  {title} {\bibinfo {title} {{Pion light-front wave
  function, parton distribution and the electromagnetic form factor}},\ }\href
  {https://doi.org/10.1088/0954-3899/42/9/095005} {\bibfield  {journal}
  {\bibinfo  {journal} {J. Phys. G}\ }\textbf {\bibinfo {volume} {42}},\
  \bibinfo {pages} {095005} (\bibinfo {year} {2015})},\ \Eprint
  {https://arxiv.org/abs/1410.6424} {arXiv:1410.6424 [hep-ph]} \BibitemShut
  {NoStop}%
\bibitem [{\citenamefont {Zhang}\ \emph {et~al.}(2017)\citenamefont {Zhang},
  \citenamefont {Radzhabov}, \citenamefont {Kochelev},\ and\ \citenamefont
  {Zhang}}]{Zhang:2017zpi}%
  \BibitemOpen
  \bibfield  {author} {\bibinfo {author} {\bibfnamefont {B.}~\bibnamefont
  {Zhang}}, \bibinfo {author} {\bibfnamefont {A.}~\bibnamefont {Radzhabov}},
  \bibinfo {author} {\bibfnamefont {N.}~\bibnamefont {Kochelev}},\ and\
  \bibinfo {author} {\bibfnamefont {P.}~\bibnamefont {Zhang}},\ }\bibfield
  {title} {\bibinfo {title} {{Pauli form factor of quark and nontrivial
  topological structure of the QCD}},\ }\href
  {https://doi.org/10.1103/PhysRevD.96.054030} {\bibfield  {journal} {\bibinfo
  {journal} {Phys. Rev. D}\ }\textbf {\bibinfo {volume} {96}},\ \bibinfo
  {pages} {054030} (\bibinfo {year} {2017})},\ \Eprint
  {https://arxiv.org/abs/1703.04333} {arXiv:1703.04333 [hep-ph]} \BibitemShut
  {NoStop}%
\bibitem [{\citenamefont {Frasca}\ and\ \citenamefont
  {Ruggieri}(2011)}]{Frasca:2011zn}%
  \BibitemOpen
  \bibfield  {author} {\bibinfo {author} {\bibfnamefont {M.}~\bibnamefont
  {Frasca}}\ and\ \bibinfo {author} {\bibfnamefont {M.}~\bibnamefont
  {Ruggieri}},\ }\bibfield  {title} {\bibinfo {title} {{Magnetic Susceptibility
  of the Quark Condensate and Polarization from Chiral Models}},\ }\href
  {https://doi.org/10.1103/PhysRevD.83.094024} {\bibfield  {journal} {\bibinfo
  {journal} {Phys. Rev. D}\ }\textbf {\bibinfo {volume} {83}},\ \bibinfo
  {pages} {094024} (\bibinfo {year} {2011})},\ \Eprint
  {https://arxiv.org/abs/1103.1194} {arXiv:1103.1194 [hep-ph]} \BibitemShut
  {NoStop}%
\bibitem [{\citenamefont {Fayazbakhsh}\ and\ \citenamefont
  {Sadooghi}(2014)}]{Fayazbakhsh:2014mca}%
  \BibitemOpen
  \bibfield  {author} {\bibinfo {author} {\bibfnamefont {S.}~\bibnamefont
  {Fayazbakhsh}}\ and\ \bibinfo {author} {\bibfnamefont {N.}~\bibnamefont
  {Sadooghi}},\ }\bibfield  {title} {\bibinfo {title} {{Anomalous magnetic
  moment of hot quarks, inverse magnetic catalysis, and reentrance of the
  chiral symmetry broken phase}},\ }\href
  {https://doi.org/10.1103/PhysRevD.90.105030} {\bibfield  {journal} {\bibinfo
  {journal} {Phys. Rev. D}\ }\textbf {\bibinfo {volume} {90}},\ \bibinfo
  {pages} {105030} (\bibinfo {year} {2014})},\ \Eprint
  {https://arxiv.org/abs/1408.5457} {arXiv:1408.5457 [hep-ph]} \BibitemShut
  {NoStop}%
\bibitem [{\citenamefont {Ferrer}\ \emph {et~al.}(2014)\citenamefont {Ferrer},
  \citenamefont {de~la Incera}, \citenamefont {Portillo},\ and\ \citenamefont
  {Quiroz}}]{Ferrer:2013noa}%
  \BibitemOpen
  \bibfield  {author} {\bibinfo {author} {\bibfnamefont {E.~J.}\ \bibnamefont
  {Ferrer}}, \bibinfo {author} {\bibfnamefont {V.}~\bibnamefont {de~la
  Incera}}, \bibinfo {author} {\bibfnamefont {I.}~\bibnamefont {Portillo}},\
  and\ \bibinfo {author} {\bibfnamefont {M.}~\bibnamefont {Quiroz}},\
  }\bibfield  {title} {\bibinfo {title} {{New look at the QCD ground state in a
  magnetic field}},\ }\href {https://doi.org/10.1103/PhysRevD.89.085034}
  {\bibfield  {journal} {\bibinfo  {journal} {Phys. Rev. D}\ }\textbf {\bibinfo
  {volume} {89}},\ \bibinfo {pages} {085034} (\bibinfo {year} {2014})},\
  \Eprint {https://arxiv.org/abs/1311.3400} {arXiv:1311.3400 [nucl-th]}
  \BibitemShut {NoStop}%
\bibitem [{\citenamefont {Tsue}\ \emph {et~al.}(2016)\citenamefont {Tsue},
  \citenamefont {da~Providencia}, \citenamefont {Providencia}, \citenamefont
  {Yamamura},\ and\ \citenamefont {Bohr}}]{Tsue:2016age}%
  \BibitemOpen
  \bibfield  {author} {\bibinfo {author} {\bibfnamefont {Y.}~\bibnamefont
  {Tsue}}, \bibinfo {author} {\bibfnamefont {J.}~\bibnamefont
  {da~Providencia}}, \bibinfo {author} {\bibfnamefont {C.}~\bibnamefont
  {Providencia}}, \bibinfo {author} {\bibfnamefont {M.}~\bibnamefont
  {Yamamura}},\ and\ \bibinfo {author} {\bibfnamefont {H.}~\bibnamefont
  {Bohr}},\ }\bibfield  {title} {\bibinfo {title} {{Spin polarization in high
  density quark matter under a strong external magnetic field}},\ }\href
  {https://doi.org/10.1142/S0218301316501068} {\bibfield  {journal} {\bibinfo
  {journal} {Int. J. Mod. Phys. E}\ }\textbf {\bibinfo {volume} {25}},\
  \bibinfo {pages} {1650106} (\bibinfo {year} {2016})},\ \Eprint
  {https://arxiv.org/abs/1607.01863} {arXiv:1607.01863 [hep-ph]} \BibitemShut
  {NoStop}%
\bibitem [{\citenamefont {Maruyama}\ \emph {et~al.}(2018)\citenamefont
  {Maruyama}, \citenamefont {Nakano}, \citenamefont {Yanase},\ and\
  \citenamefont {Yoshinaga}}]{Maruyama:2018fpl}%
  \BibitemOpen
  \bibfield  {author} {\bibinfo {author} {\bibfnamefont {T.}~\bibnamefont
  {Maruyama}}, \bibinfo {author} {\bibfnamefont {E.}~\bibnamefont {Nakano}},
  \bibinfo {author} {\bibfnamefont {K.}~\bibnamefont {Yanase}},\ and\ \bibinfo
  {author} {\bibfnamefont {N.}~\bibnamefont {Yoshinaga}},\ }\bibfield  {title}
  {\bibinfo {title} {{Spin polarized phases in strongly interacting matter:
  interplay between axial-vector and tensor mean fields}},\ }\href
  {https://doi.org/10.1103/PhysRevD.97.114014} {\bibfield  {journal} {\bibinfo
  {journal} {Phys. Rev. D}\ }\textbf {\bibinfo {volume} {97}},\ \bibinfo
  {pages} {114014} (\bibinfo {year} {2018})},\ \Eprint
  {https://arxiv.org/abs/1803.08315} {arXiv:1803.08315 [hep-ph]} \BibitemShut
  {NoStop}%
\bibitem [{\citenamefont {Chaudhuri}\ \emph {et~al.}(2019)\citenamefont
  {Chaudhuri}, \citenamefont {Ghosh}, \citenamefont {Sarkar},\ and\
  \citenamefont {Roy}}]{Chaudhuri:2019lbw}%
  \BibitemOpen
  \bibfield  {author} {\bibinfo {author} {\bibfnamefont {N.}~\bibnamefont
  {Chaudhuri}}, \bibinfo {author} {\bibfnamefont {S.}~\bibnamefont {Ghosh}},
  \bibinfo {author} {\bibfnamefont {S.}~\bibnamefont {Sarkar}},\ and\ \bibinfo
  {author} {\bibfnamefont {P.}~\bibnamefont {Roy}},\ }\bibfield  {title}
  {\bibinfo {title} {{Effect of the anomalous magnetic moment of quarks on the
  phase structure and mesonic properties in the NJL model}},\ }\href
  {https://doi.org/10.1103/PhysRevD.99.116025} {\bibfield  {journal} {\bibinfo
  {journal} {Phys. Rev. D}\ }\textbf {\bibinfo {volume} {99}},\ \bibinfo
  {pages} {116025} (\bibinfo {year} {2019})},\ \Eprint
  {https://arxiv.org/abs/1907.03990} {arXiv:1907.03990 [nucl-th]} \BibitemShut
  {NoStop}%
\bibitem [{\citenamefont {Mei}\ and\ \citenamefont {Mao}(2020)}]{Mei:2020jzn}%
  \BibitemOpen
  \bibfield  {author} {\bibinfo {author} {\bibfnamefont {J.}~\bibnamefont
  {Mei}}\ and\ \bibinfo {author} {\bibfnamefont {S.}~\bibnamefont {Mao}},\
  }\bibfield  {title} {\bibinfo {title} {{Inverse catalysis effect of the quark
  anomalous magnetic moment to chiral restoration and deconfinement phase
  transitions}},\ }\href {https://doi.org/10.1103/PhysRevD.102.114035}
  {\bibfield  {journal} {\bibinfo  {journal} {Phys. Rev. D}\ }\textbf {\bibinfo
  {volume} {102}},\ \bibinfo {pages} {114035} (\bibinfo {year} {2020})},\
  \Eprint {https://arxiv.org/abs/2008.12123} {arXiv:2008.12123 [hep-ph]}
  \BibitemShut {NoStop}%
\bibitem [{\citenamefont {Xu}\ \emph {et~al.}(2021)\citenamefont {Xu},
  \citenamefont {Chao},\ and\ \citenamefont {Huang}}]{Xu:2020yag}%
  \BibitemOpen
  \bibfield  {author} {\bibinfo {author} {\bibfnamefont {K.}~\bibnamefont
  {Xu}}, \bibinfo {author} {\bibfnamefont {J.}~\bibnamefont {Chao}},\ and\
  \bibinfo {author} {\bibfnamefont {M.}~\bibnamefont {Huang}},\ }\bibfield
  {title} {\bibinfo {title} {{Effect of the anomalous magnetic moment of quarks
  on magnetized QCD matter and meson spectra}},\ }\href
  {https://doi.org/10.1103/PhysRevD.103.076015} {\bibfield  {journal} {\bibinfo
   {journal} {Phys. Rev. D}\ }\textbf {\bibinfo {volume} {103}},\ \bibinfo
  {pages} {076015} (\bibinfo {year} {2021})},\ \Eprint
  {https://arxiv.org/abs/2007.13122} {arXiv:2007.13122 [hep-ph]} \BibitemShut
  {NoStop}%
\bibitem [{\citenamefont {Aguirre}(2021)}]{Aguirre:2021ljk}%
  \BibitemOpen
  \bibfield  {author} {\bibinfo {author} {\bibfnamefont {R.~M.}\ \bibnamefont
  {Aguirre}},\ }\bibfield  {title} {\bibinfo {title} {{Effects of the anomalous
  magnetic moments of the quarks on the neutral pion properties within a SU(2)
  Nambu\textendash{}Jona Lasinio model}},\ }\href
  {https://doi.org/10.1140/epja/s10050-021-00480-1} {\bibfield  {journal}
  {\bibinfo  {journal} {Eur. Phys. J. A}\ }\textbf {\bibinfo {volume} {57}},\
  \bibinfo {pages} {166} (\bibinfo {year} {2021})}\BibitemShut {NoStop}%
\bibitem [{\citenamefont {Coleman}(1973)}]{Coleman:1973ci}%
  \BibitemOpen
  \bibfield  {author} {\bibinfo {author} {\bibfnamefont {S.~R.}\ \bibnamefont
  {Coleman}},\ }\bibfield  {title} {\bibinfo {title} {{There are no Goldstone
  bosons in two-dimensions}},\ }\href {https://doi.org/10.1007/BF01646487}
  {\bibfield  {journal} {\bibinfo  {journal} {Commun. Math. Phys.}\ }\textbf
  {\bibinfo {volume} {31}},\ \bibinfo {pages} {259} (\bibinfo {year}
  {1973})}\BibitemShut {NoStop}%
\bibitem [{\citenamefont {Gusynin}\ \emph {et~al.}(1996)\citenamefont
  {Gusynin}, \citenamefont {Miransky},\ and\ \citenamefont
  {Shovkovy}}]{Gusynin:1995nb}%
  \BibitemOpen
  \bibfield  {author} {\bibinfo {author} {\bibfnamefont {V.~P.}\ \bibnamefont
  {Gusynin}}, \bibinfo {author} {\bibfnamefont {V.~A.}\ \bibnamefont
  {Miransky}},\ and\ \bibinfo {author} {\bibfnamefont {I.~A.}\ \bibnamefont
  {Shovkovy}},\ }\bibfield  {title} {\bibinfo {title} {{Dimensional reduction
  and catalysis of dynamical symmetry breaking by a magnetic field}},\ }\href
  {https://doi.org/10.1016/0550-3213(96)00021-1} {\bibfield  {journal}
  {\bibinfo  {journal} {Nucl. Phys. B}\ }\textbf {\bibinfo {volume} {462}},\
  \bibinfo {pages} {249} (\bibinfo {year} {1996})},\ \Eprint
  {https://arxiv.org/abs/hep-ph/9509320} {arXiv:hep-ph/9509320} \BibitemShut
  {NoStop}%
\bibitem [{\citenamefont {Agasian}\ and\ \citenamefont
  {Shushpanov}(2001)}]{Agasian:2001ym}%
  \BibitemOpen
  \bibfield  {author} {\bibinfo {author} {\bibfnamefont {N.~O.}\ \bibnamefont
  {Agasian}}\ and\ \bibinfo {author} {\bibfnamefont {I.~A.}\ \bibnamefont
  {Shushpanov}},\ }\bibfield  {title} {\bibinfo {title}
  {{Gell-Mann-Oakes-Renner relation in a magnetic field at finite
  temperature}},\ }\href {https://doi.org/10.1088/1126-6708/2001/10/006}
  {\bibfield  {journal} {\bibinfo  {journal} {JHEP}\ }\textbf {\bibinfo
  {volume} {10}},\ \bibinfo {pages} {006}},\ \Eprint
  {https://arxiv.org/abs/hep-ph/0107128} {arXiv:hep-ph/0107128} \BibitemShut
  {NoStop}%
\bibitem [{\citenamefont {Avancini}\ \emph {et~al.}(2016)\citenamefont
  {Avancini}, \citenamefont {Tavares},\ and\ \citenamefont
  {Pinto}}]{Avancini:2015ady}%
  \BibitemOpen
  \bibfield  {author} {\bibinfo {author} {\bibfnamefont {S.~S.}\ \bibnamefont
  {Avancini}}, \bibinfo {author} {\bibfnamefont {W.~R.}\ \bibnamefont
  {Tavares}},\ and\ \bibinfo {author} {\bibfnamefont {M.~B.}\ \bibnamefont
  {Pinto}},\ }\bibfield  {title} {\bibinfo {title} {{Properties of magnetized
  neutral mesons within a full RPA evaluation}},\ }\href
  {https://doi.org/10.1103/PhysRevD.93.014010} {\bibfield  {journal} {\bibinfo
  {journal} {Phys. Rev. D}\ }\textbf {\bibinfo {volume} {93}},\ \bibinfo
  {pages} {014010} (\bibinfo {year} {2016})},\ \Eprint
  {https://arxiv.org/abs/1511.06261} {arXiv:1511.06261 [hep-ph]} \BibitemShut
  {NoStop}%
\bibitem [{\citenamefont {Yamamoto}\ \emph {et~al.}(2007)\citenamefont
  {Yamamoto}, \citenamefont {Tachibana}, \citenamefont {Hatsuda},\ and\
  \citenamefont {Baym}}]{Yamamoto:2007ah}%
  \BibitemOpen
  \bibfield  {author} {\bibinfo {author} {\bibfnamefont {N.}~\bibnamefont
  {Yamamoto}}, \bibinfo {author} {\bibfnamefont {M.}~\bibnamefont {Tachibana}},
  \bibinfo {author} {\bibfnamefont {T.}~\bibnamefont {Hatsuda}},\ and\ \bibinfo
  {author} {\bibfnamefont {G.}~\bibnamefont {Baym}},\ }\bibfield  {title}
  {\bibinfo {title} {{Phase structure, collective modes, and the axial anomaly
  in dense QCD}},\ }\href {https://doi.org/10.1103/PhysRevD.76.074001}
  {\bibfield  {journal} {\bibinfo  {journal} {Phys. Rev. D}\ }\textbf {\bibinfo
  {volume} {76}},\ \bibinfo {pages} {074001} (\bibinfo {year} {2007})},\
  \Eprint {https://arxiv.org/abs/0704.2654} {arXiv:0704.2654 [hep-ph]}
  \BibitemShut {NoStop}%
\bibitem [{\citenamefont {Song}\ and\ \citenamefont
  {Baym}(2017)}]{Song:2017dws}%
  \BibitemOpen
  \bibfield  {author} {\bibinfo {author} {\bibfnamefont {Y.}~\bibnamefont
  {Song}}\ and\ \bibinfo {author} {\bibfnamefont {G.}~\bibnamefont {Baym}},\
  }\bibfield  {title} {\bibinfo {title} {{Generalized Nambu-Goldstone pion in
  dense matter: A schematic Nambu\textendash{}Jona-Lasinio model}},\ }\href
  {https://doi.org/10.1103/PhysRevC.96.025206} {\bibfield  {journal} {\bibinfo
  {journal} {Phys. Rev. C}\ }\textbf {\bibinfo {volume} {96}},\ \bibinfo
  {pages} {025206} (\bibinfo {year} {2017})},\ \Eprint
  {https://arxiv.org/abs/1703.08236} {arXiv:1703.08236 [nucl-th]} \BibitemShut
  {NoStop}%
\bibitem [{\citenamefont {Dittrich}\ and\ \citenamefont
  {Gies}(2000)}]{Dittrich:2000zu}%
  \BibitemOpen
  \bibfield  {author} {\bibinfo {author} {\bibfnamefont {W.}~\bibnamefont
  {Dittrich}}\ and\ \bibinfo {author} {\bibfnamefont {H.}~\bibnamefont
  {Gies}},\ }\href {https://doi.org/10.1007/3-540-45585-X} {\emph {\bibinfo
  {title} {{Probing the quantum vacuum. Perturbative effective action approach
  in quantum electrodynamics and its application}}}},\ Vol.\ \bibinfo {volume}
  {166}\ (\bibinfo  {publisher} {Springer Tracts in Modern Physics},\ \bibinfo
  {year} {2000})\BibitemShut {NoStop}%
\bibitem [{\citenamefont {Schwinger}(1951)}]{Schwinger:1951nm}%
  \BibitemOpen
  \bibfield  {author} {\bibinfo {author} {\bibfnamefont {J.~S.}\ \bibnamefont
  {Schwinger}},\ }\bibfield  {title} {\bibinfo {title} {{On gauge invariance
  and vacuum polarization}},\ }\href {https://doi.org/10.1103/PhysRev.82.664}
  {\bibfield  {journal} {\bibinfo  {journal} {Phys. Rev.}\ }\textbf {\bibinfo
  {volume} {82}},\ \bibinfo {pages} {664} (\bibinfo {year} {1951})}\BibitemShut
  {NoStop}%
\bibitem [{\citenamefont {Miransky}\ and\ \citenamefont
  {Shovkovy}(2015)}]{Miransky:2015ava}%
  \BibitemOpen
  \bibfield  {author} {\bibinfo {author} {\bibfnamefont {V.~A.}\ \bibnamefont
  {Miransky}}\ and\ \bibinfo {author} {\bibfnamefont {I.~A.}\ \bibnamefont
  {Shovkovy}},\ }\bibfield  {title} {\bibinfo {title} {{Quantum field theory in
  a magnetic field: From quantum chromodynamics to graphene and Dirac
  semimetals}},\ }\href {https://doi.org/10.1016/j.physrep.2015.02.003}
  {\bibfield  {journal} {\bibinfo  {journal} {Phys. Rept.}\ }\textbf {\bibinfo
  {volume} {576}},\ \bibinfo {pages} {1} (\bibinfo {year} {2015})},\ \Eprint
  {https://arxiv.org/abs/1503.00732} {arXiv:1503.00732 [hep-ph]} \BibitemShut
  {NoStop}%
\bibitem [{\citenamefont {Pisarski}\ \emph {et~al.}(2020)\citenamefont
  {Pisarski}, \citenamefont {Tsvelik},\ and\ \citenamefont
  {Valgushev}}]{Pisarski:2020dnx}%
  \BibitemOpen
  \bibfield  {author} {\bibinfo {author} {\bibfnamefont {R.~D.}\ \bibnamefont
  {Pisarski}}, \bibinfo {author} {\bibfnamefont {A.~M.}\ \bibnamefont
  {Tsvelik}},\ and\ \bibinfo {author} {\bibfnamefont {S.}~\bibnamefont
  {Valgushev}},\ }\bibfield  {title} {\bibinfo {title} {{How transverse thermal
  fluctuations disorder a condensate of chiral spirals into a quantum spin
  liquid}},\ }\href {https://doi.org/10.1103/PhysRevD.102.016015} {\bibfield
  {journal} {\bibinfo  {journal} {Phys. Rev. D}\ }\textbf {\bibinfo {volume}
  {102}},\ \bibinfo {pages} {016015} (\bibinfo {year} {2020})},\ \Eprint
  {https://arxiv.org/abs/2005.10259} {arXiv:2005.10259 [hep-ph]} \BibitemShut
  {NoStop}%
\bibitem [{\citenamefont {He}\ \emph {et~al.}(2005)\citenamefont {He},
  \citenamefont {Jin},\ and\ \citenamefont {Zhuang}}]{He:2005nk}%
  \BibitemOpen
  \bibfield  {author} {\bibinfo {author} {\bibfnamefont {L.-y.}\ \bibnamefont
  {He}}, \bibinfo {author} {\bibfnamefont {M.}~\bibnamefont {Jin}},\ and\
  \bibinfo {author} {\bibfnamefont {P.-f.}\ \bibnamefont {Zhuang}},\ }\bibfield
   {title} {\bibinfo {title} {{Pion superfluidity and meson properties at
  finite isospin density}},\ }\href
  {https://doi.org/10.1103/PhysRevD.71.116001} {\bibfield  {journal} {\bibinfo
  {journal} {Phys. Rev. D}\ }\textbf {\bibinfo {volume} {71}},\ \bibinfo
  {pages} {116001} (\bibinfo {year} {2005})},\ \Eprint
  {https://arxiv.org/abs/hep-ph/0503272} {arXiv:hep-ph/0503272} \BibitemShut
  {NoStop}%
\bibitem [{\citenamefont {Ferrer}\ and\ \citenamefont {de~la
  Incera}(2018)}]{Ferrer:2015iop}%
  \BibitemOpen
  \bibfield  {author} {\bibinfo {author} {\bibfnamefont {E.~J.}\ \bibnamefont
  {Ferrer}}\ and\ \bibinfo {author} {\bibfnamefont {V.}~\bibnamefont {de~la
  Incera}},\ }\bibfield  {title} {\bibinfo {title} {{Novel Topological Effects
  in Dense QCD in a Magnetic Field}},\ }\href
  {https://doi.org/10.1016/j.nuclphysb.2018.04.009} {\bibfield  {journal}
  {\bibinfo  {journal} {Nucl. Phys. B}\ }\textbf {\bibinfo {volume} {931}},\
  \bibinfo {pages} {192} (\bibinfo {year} {2018})},\ \Eprint
  {https://arxiv.org/abs/1512.03972} {arXiv:1512.03972 [nucl-th]} \BibitemShut
  {NoStop}%
\bibitem [{\citenamefont {Brauner}\ and\ \citenamefont
  {Yamamoto}(2017)}]{Brauner:2016pko}%
  \BibitemOpen
  \bibfield  {author} {\bibinfo {author} {\bibfnamefont {T.}~\bibnamefont
  {Brauner}}\ and\ \bibinfo {author} {\bibfnamefont {N.}~\bibnamefont
  {Yamamoto}},\ }\bibfield  {title} {\bibinfo {title} {{Chiral Soliton Lattice
  and Charged Pion Condensation in Strong Magnetic Fields}},\ }\href
  {https://doi.org/10.1007/JHEP04(2017)132} {\bibfield  {journal} {\bibinfo
  {journal} {JHEP}\ }\textbf {\bibinfo {volume} {04}},\ \bibinfo {pages}
  {132}},\ \Eprint {https://arxiv.org/abs/1609.05213} {arXiv:1609.05213
  [hep-ph]} \BibitemShut {NoStop}%
\bibitem [{\citenamefont {Chao}\ \emph {et~al.}(2020)\citenamefont {Chao},
  \citenamefont {Huang},\ and\ \citenamefont {Radzhabov}}]{Chao:2018ejd}%
  \BibitemOpen
  \bibfield  {author} {\bibinfo {author} {\bibfnamefont {J.}~\bibnamefont
  {Chao}}, \bibinfo {author} {\bibfnamefont {M.}~\bibnamefont {Huang}},\ and\
  \bibinfo {author} {\bibfnamefont {A.}~\bibnamefont {Radzhabov}},\ }\bibfield
  {title} {\bibinfo {title} {{Charged pion condensation in anti-parallel
  electromagnetic fields and nonzero isospin density}},\ }\href
  {https://doi.org/10.1088/1674-1137/44/3/034105} {\bibfield  {journal}
  {\bibinfo  {journal} {Chin. Phys. C}\ }\textbf {\bibinfo {volume} {44}},\
  \bibinfo {pages} {034105} (\bibinfo {year} {2020})},\ \Eprint
  {https://arxiv.org/abs/1805.00614} {arXiv:1805.00614 [hep-ph]} \BibitemShut
  {NoStop}%
\bibitem [{\citenamefont {Vovchenko}\ \emph {et~al.}(2021)\citenamefont
  {Vovchenko}, \citenamefont {Brandt}, \citenamefont {Cuteri}, \citenamefont
  {Endr\H{o}di}, \citenamefont {Hajkarim},\ and\ \citenamefont
  {Schaffner-Bielich}}]{Vovchenko:2020crk}%
  \BibitemOpen
  \bibfield  {author} {\bibinfo {author} {\bibfnamefont {V.}~\bibnamefont
  {Vovchenko}}, \bibinfo {author} {\bibfnamefont {B.~B.}\ \bibnamefont
  {Brandt}}, \bibinfo {author} {\bibfnamefont {F.}~\bibnamefont {Cuteri}},
  \bibinfo {author} {\bibfnamefont {G.}~\bibnamefont {Endr\H{o}di}}, \bibinfo
  {author} {\bibfnamefont {F.}~\bibnamefont {Hajkarim}},\ and\ \bibinfo
  {author} {\bibfnamefont {J.}~\bibnamefont {Schaffner-Bielich}},\ }\bibfield
  {title} {\bibinfo {title} {{Pion Condensation in the Early Universe at
  Nonvanishing Lepton Flavor Asymmetry and Its Gravitational Wave
  Signatures}},\ }\href {https://doi.org/10.1103/PhysRevLett.126.012701}
  {\bibfield  {journal} {\bibinfo  {journal} {Phys. Rev. Lett.}\ }\textbf
  {\bibinfo {volume} {126}},\ \bibinfo {pages} {012701} (\bibinfo {year}
  {2021})},\ \Eprint {https://arxiv.org/abs/2009.02309} {arXiv:2009.02309
  [hep-ph]} \BibitemShut {NoStop}%
\bibitem [{\citenamefont {Chernodub}(2011)}]{Chernodub:2011mc}%
  \BibitemOpen
  \bibfield  {author} {\bibinfo {author} {\bibfnamefont {M.~N.}\ \bibnamefont
  {Chernodub}},\ }\bibfield  {title} {\bibinfo {title} {{Spontaneous
  electromagnetic superconductivity of vacuum in strong magnetic field:
  evidence from the Nambu--Jona-Lasinio model}},\ }\href
  {https://doi.org/10.1103/PhysRevLett.106.142003} {\bibfield  {journal}
  {\bibinfo  {journal} {Phys. Rev. Lett.}\ }\textbf {\bibinfo {volume} {106}},\
  \bibinfo {pages} {142003} (\bibinfo {year} {2011})},\ \Eprint
  {https://arxiv.org/abs/1101.0117} {arXiv:1101.0117 [hep-ph]} \BibitemShut
  {NoStop}%
\bibitem [{\citenamefont {Hidaka}\ and\ \citenamefont
  {Yamamoto}(2013)}]{Hidaka:2012mz}%
  \BibitemOpen
  \bibfield  {author} {\bibinfo {author} {\bibfnamefont {Y.}~\bibnamefont
  {Hidaka}}\ and\ \bibinfo {author} {\bibfnamefont {A.}~\bibnamefont
  {Yamamoto}},\ }\bibfield  {title} {\bibinfo {title} {{Charged vector mesons
  in a strong magnetic field}},\ }\href
  {https://doi.org/10.1103/PhysRevD.87.094502} {\bibfield  {journal} {\bibinfo
  {journal} {Phys. Rev. D}\ }\textbf {\bibinfo {volume} {87}},\ \bibinfo
  {pages} {094502} (\bibinfo {year} {2013})},\ \Eprint
  {https://arxiv.org/abs/1209.0007} {arXiv:1209.0007 [hep-ph]} \BibitemShut
  {NoStop}%
\bibitem [{\citenamefont {Hofmann}(2021)}]{Hofmann:2021bac}%
  \BibitemOpen
  \bibfield  {author} {\bibinfo {author} {\bibfnamefont {C.~P.}\ \bibnamefont
  {Hofmann}},\ }\bibfield  {title} {\bibinfo {title} {{Diamagnetic and
  paramagnetic phases in low-energy quantum chromodynamics}},\ }\href
  {https://doi.org/10.1016/j.physletb.2021.136384} {\bibfield  {journal}
  {\bibinfo  {journal} {Phys. Lett. B}\ }\textbf {\bibinfo {volume} {818}},\
  \bibinfo {pages} {136384} (\bibinfo {year} {2021})},\ \Eprint
  {https://arxiv.org/abs/2103.04937} {arXiv:2103.04937 [hep-ph]} \BibitemShut
  {NoStop}%
\bibitem [{\citenamefont {D'Elia}\ \emph {et~al.}(2018)\citenamefont {D'Elia},
  \citenamefont {Manigrasso}, \citenamefont {Negro},\ and\ \citenamefont
  {Sanfilippo}}]{DElia:2018xwo}%
  \BibitemOpen
  \bibfield  {author} {\bibinfo {author} {\bibfnamefont {M.}~\bibnamefont
  {D'Elia}}, \bibinfo {author} {\bibfnamefont {F.}~\bibnamefont {Manigrasso}},
  \bibinfo {author} {\bibfnamefont {F.}~\bibnamefont {Negro}},\ and\ \bibinfo
  {author} {\bibfnamefont {F.}~\bibnamefont {Sanfilippo}},\ }\bibfield  {title}
  {\bibinfo {title} {{QCD phase diagram in a magnetic background for different
  values of the pion mass}},\ }\href
  {https://doi.org/10.1103/PhysRevD.98.054509} {\bibfield  {journal} {\bibinfo
  {journal} {Phys. Rev. D}\ }\textbf {\bibinfo {volume} {98}},\ \bibinfo
  {pages} {054509} (\bibinfo {year} {2018})},\ \Eprint
  {https://arxiv.org/abs/1808.07008} {arXiv:1808.07008 [hep-lat]} \BibitemShut
  {NoStop}%
\bibitem [{\citenamefont {Endrődi}\ \emph {et~al.}(2019)\citenamefont
  {Endrődi}, \citenamefont {Giordano}, \citenamefont {Katz}, \citenamefont
  {Kovács},\ and\ \citenamefont {Pittler}}]{Endrodi:2019zrl}%
  \BibitemOpen
  \bibfield  {author} {\bibinfo {author} {\bibfnamefont {G.}~\bibnamefont
  {Endrődi}}, \bibinfo {author} {\bibfnamefont {M.}~\bibnamefont {Giordano}},
  \bibinfo {author} {\bibfnamefont {S.~D.}\ \bibnamefont {Katz}}, \bibinfo
  {author} {\bibfnamefont {T.~G.}\ \bibnamefont {Kovács}},\ and\ \bibinfo
  {author} {\bibfnamefont {F.}~\bibnamefont {Pittler}},\ }\bibfield  {title}
  {\bibinfo {title} {{Magnetic catalysis and inverse catalysis for heavy
  pions}},\ }\href {https://doi.org/10.1007/JHEP07(2019)007} {\bibfield
  {journal} {\bibinfo  {journal} {JHEP}\ }\textbf {\bibinfo {volume} {07}},\
  \bibinfo {pages} {007}},\ \Eprint {https://arxiv.org/abs/1904.10296}
  {arXiv:1904.10296 [hep-lat]} \BibitemShut {NoStop}%
\bibitem [{\citenamefont {Lajer}\ \emph {et~al.}(2022)\citenamefont {Lajer},
  \citenamefont {Konik}, \citenamefont {Pisarski},\ and\ \citenamefont
  {Tsvelik}}]{Lajer:2021kcz}%
  \BibitemOpen
  \bibfield  {author} {\bibinfo {author} {\bibfnamefont {M.}~\bibnamefont
  {Lajer}}, \bibinfo {author} {\bibfnamefont {R.~M.}\ \bibnamefont {Konik}},
  \bibinfo {author} {\bibfnamefont {R.~D.}\ \bibnamefont {Pisarski}},\ and\
  \bibinfo {author} {\bibfnamefont {A.~M.}\ \bibnamefont {Tsvelik}},\
  }\bibfield  {title} {\bibinfo {title} {{When cold, dense quarks in 1+1 and
  3+1 dimensions are not a Fermi liquid}},\ }\href
  {https://doi.org/10.1103/PhysRevD.105.054035} {\bibfield  {journal} {\bibinfo
   {journal} {Phys. Rev. D}\ }\textbf {\bibinfo {volume} {105}},\ \bibinfo
  {pages} {054035} (\bibinfo {year} {2022})},\ \Eprint
  {https://arxiv.org/abs/2112.10238} {arXiv:2112.10238 [hep-th]} \BibitemShut
  {NoStop}%
\end{thebibliography}%
\end{document}